\documentclass[aps,prl,reprint,superscriptaddress]{revtex4-2}

\usepackage{graphicx}
\usepackage{float}
\makeatletter
\let\newfloat\newfloat@ltx
\makeatother
\usepackage{subfig}
\usepackage{dcolumn}
\usepackage{lmodern}
\usepackage{bm}

\usepackage[utf8]{inputenc}
\usepackage[T1]{fontenc}
\usepackage{booktabs, array, mathptmx, tabularx, booktabs, lipsum, amsmath,multirow}
\usepackage{siunitx, xcolor}
\usepackage[version=4]{mhchem}
\usepackage{changes}

\usepackage{braket}
\usepackage{algorithm}
\usepackage{algorithmicx}
\usepackage[noend]{algpseudocode}
\usepackage{microtype}
\usepackage[colorlinks,linkcolor=blue,anchorcolor=blue,citecolor=blue]{hyperref}

\begin{document}


\title{Long-algorithm based quantum search for gravitational wave}


\author{Fangzhou Guo}
\email[]{guofangzhou23@mails.ucas.ac.cn}
\affiliation{School of Fundamental Physics and Mathematical Sciences, Hangzhou Institute for Advanced Study, UCAS, Hangzhou 310024, China}
\affiliation{University of Chinese Academy of Sciences, Beijing 100049, China}
\affiliation{Institute of Theoretical Physics, Chinese Academy of Sciences, Beijing 100190, China}
\affiliation{International Centre for Theoretical Physics Asia-Pacific, UCAS, Beijing 100190, China}
\author{Jibo He}
\email[]{jibo.he@ucas.ac.cn}
\affiliation{School of Fundamental Physics and Mathematical Sciences, Hangzhou Institute for Advanced Study, UCAS, Hangzhou 310024, China}
\affiliation{University of Chinese Academy of Sciences, Beijing 100049, China}
\affiliation{International Centre for Theoretical Physics Asia-Pacific, UCAS, Beijing 100190, China}
\affiliation{Taiji Laboratory for Gravitational Wave Universe (Beijing/Hangzhou), UCAS, Beijing 100190, China}

\date{\today}

\begin{abstract}
Gravitational wave astronomy is rapidly advancing with the development of new observatories, leading to an increasing volume and complexity of data. This trend places growing pressure on classical data analysis methods and motivates the exploration of quantum approaches. In this work, we introduce a quantum matched filtering framework for gravitational-wave detection based on the Long algorithm, marking its first application to the gravitational-wave data analysis. Numerical simulations show that the proposed approach preserves the quadratic speedup of quantum search while exhibiting significantly improved robustness, thereby overcoming key limitations of 
the Grover-algorithm based methods. 

\end{abstract}


\maketitle


\textit{Introduction\textemdash}Gravitational wave astronomy has advanced rapidly since the first detection of the compact binary coalescence GW150914~\cite{abbott2016observation}, opening a new observational window on the Universe. Ground-based detectors such as Advanced LIGO~\cite{aasi2015advanced}, Advanced Virgo~\cite{acernese2014advanced} and KAGRA~\cite{somiya2012detector} have already revealed a rich population of compact binaries, and further improvements in sensitivity are expected to significantly increase detection rates. Ongoing upgrades and planned improvements, together with future third generation ground-based detectors such as the Einstein Telescope~\cite{punturo2010einstein} and Cosmic Explorer~\cite{reitze2019cosmic}, aim to further enhance detector sensitivity, enabling the detection of weaker and more distant sources. While this progress significantly increases the scientific reach of ground-based observatories, it also places growing demands on data analysis pipelines, as higher sensitivity leads to increased detection rates and a larger volume of data that must be analyzed with low latency to enable the multi-messenger observation. 

At the same time, future space-based missions including LISA~\cite{amaro2017laser}, Taiji~\cite{hu2017taiji}, and TianQin~\cite{luo2016tianqin} will target the millihertz band. In this regime, the target sources such as massive black hole binaries (MBHBs), extreme mass-ratio inspirals (EMRIs), and Galactic compact binaries (GCBs) produce long duration signals characterized by higher dimensional parameter spaces, which increases the computational cost of data analysis.

These trends highlight a common challenge across current and future gravitational wave data analysis: the rapidly increasing computational cost of signal searches driven by improved sensitivity and expanding parameter spaces. This motivates the exploration of new data analysis methods that can surpass classical computational scaling, including quantum approaches.

For both ground-based and space-based detectors, matched filtering~\cite{owen1999matched} remains the primary detection strategy. In this approach, the detector data are correlated against a bank of template waveforms constructed from signals with different source parameters. Because these parameters are not known a priori, the template bank must densely cover a high-dimensional astrophysical parameter space, and the computational cost of matched filtering scales linearly with the number of templates.

For ground-based detectors, this cost is compounded by the requirement of low latency analysis. Rapid identification of candidate events is essential for enabling timely multi-messenger follow up observations, placing stringent constraints on the speed of matched-filtering pipelines even when the total data volume is moderate. For space-based observations, the challenge is further amplified: signals from sources such as massive black hole binaries can last from months to years, and are described by higher dimensional parameter spaces. The combination of long duration signals and large template banks leads to a substantial increase in computational cost, making matched filtering particularly demanding in the millihertz band. These challenges expose the limitations of classical matched filtering and motivate the development of quantum-enhanced matched filtering techniques.

Grover’s algorithm~\cite{grover1996fast} is a quantum search algorithm that provides a quadratic speedup for searching unstructured databases. By exploiting quantum superposition and interference, Grover’s algorithm reduces the computational complexity of locating marked items from $O(N)$ in classical exhaustive search to $O(\sqrt{N})$, where $N$ is the size of the database. This speedup is achieved through an iterative amplitude amplification process, in which probability amplitude is coherently transferred from non-solution states to the target subspace. Since many scientific data analysis tasks can be naturally formulated as search or optimization problems over large, unstructured parameter spaces, Grover’s algorithm has attracted significant interest as a potential route to accelerating classically expensive searches. In particular, proposed applications include pattern matching~\cite{tezuka2022grover}, combinatorial optimization~\cite{durr1996quantum} and data analysis tasks in high energy physics~\cite{wei2020quantum}. Motivated by the quadratic speedup offered by Grover’s algorithm, a quantum matched filtering (QMF) schemes have been proposed to accelerate gravitational wave searches, suggesting the possibility of substantial reductions in computational complexity theoretically~\cite{gao2022quantum}.

However, the standard Grover algorithm is probabilistic in nature, its success probability is generally below unity. The Long algorithm~\cite{long2001grover} is a modified version of Grover’s algorithm that achieves unit success probability, making it naturally suited to gravitational wave searches where robustness against mismatch is essential. 

In this work, we introduce a quantum search method based on the Long algorithm and quantum counting. Through numerical simulations, we evaluate the detection performance and computational cost of this method and compare it with the original Grover-based quantum matched filtering (QMF) approach as proposed in Ref.~\cite{gao2022quantum} using the GW150914 event and simulated massive black hole binary signals. It is found that the Long-based approach preserves the quadratic speedup of quantum search while exhibiting significantly improved robustness in search scenarios, thereby overcoming the key limitations of the Grover-based approaches as reported in our previous work~\cite{guo2025quantum}.

 
\textit{Error analysis of Grover-based and Long-based algorithm\textemdash}
In classical gravitational wave data analysis, matched filtering searches for signals by correlating detector data with a bank of template waveforms. In a Grover-based QMF framework, the template index is encoded into a quantum register, and the search over the template bank is mapped onto a quantum amplitude amplification problem.

The algorithm begins by preparing a uniform superposition over all basis states:
\begin{equation}
\ket{s} = \frac{1}{\sqrt{N}} \sum_{i=0}^{N-1} \ket{i}.
\end{equation}
The first key component of Grover's algorithm is the \textit{oracle} operator $U_f$, which encodes the function $f(x)$ by flipping the sign of the marked states:
\begin{equation} \label{kick}
U_f \ket{x} = (-1)^{f(x)} \ket{x}.
\end{equation}
This means that if $x$ is a solution, the amplitude of $\ket{x}$ is inverted in phase, while all other basis states remain unchanged.

Geometrically, Grover’s algorithm performs a rotation in the two-dimensional subspace spanned by $\ket{\alpha}$ (the equal superposition of unmarked states) and $\ket{\beta}$ (the equal superposition of marked states). The initial state $\ket{s}$ lies at an angle $\theta$ from $\ket{\alpha}$, where
\begin{equation}\label{theta_r}
\sin(\theta) = \sqrt{\frac{r}{N}}.
\end{equation}
where $r$ is the number of target templates, and $N$ is the total number of templates. Each Grover iteration consists of applying the oracle $U_f$ followed by the Grover diffusion operator $D = 2\ket{s}\bra{s} - I$. Together, they form the Grover operator
\begin{equation}
G = D \cdot U_f,
\end{equation}
which rotates the state vector by an angle of $2\theta$ toward $\ket{\beta}$.
After approximately
\begin{equation} \label{optk_appro}
k \approx \left\lfloor \frac{\pi/2-\theta}{2\theta}  \right\rfloor \approx \left\lfloor \frac{\pi}{4} \sqrt{\frac{N}{r}} - \frac{1}{2}\right\rfloor
\end{equation}
iterations, the state vector is close to $\ket{\beta}$, and a measurement yields a correct result with high probability, completing the quantum search in $O(\sqrt{N})$ steps.

In realistic gravitational wave searches, the effective rotation angle $2\theta$ of the Grover operator is generally unknown. To address this issue, quantum counting has been proposed to estimate $\theta$ by performing phase estimation on the Grover operator $G$. In the two-dimensional subspace spanned by $\ket{\alpha}$ and $\ket{\beta}$, the operator $G$ has eigenvalues $e^{\pm i 2\theta}$, which allows $\theta$ to be inferred from the estimated eigenphases.

However, quantum counting provides only an approximate estimate of $\theta$, with a finite statistical uncertainty determined by the number of ancilla qubits and the number of controlled-$G$ operations used in the phase-estimation procedure. This uncertainty directly affects the determination of the optimal number of Grover iterations. Since Grover’s algorithm relies on amplitude amplification, successive applications of $G$ rotate the quantum state toward the target subspace by a fixed angle $2\theta$. An inaccurate estimate of $\theta$ therefore leads to a mismatch between the estimated and true optimal iteration counts.

\begin{figure}[htbp]      
    \centering             
    \includegraphics[width=0.4\textwidth]{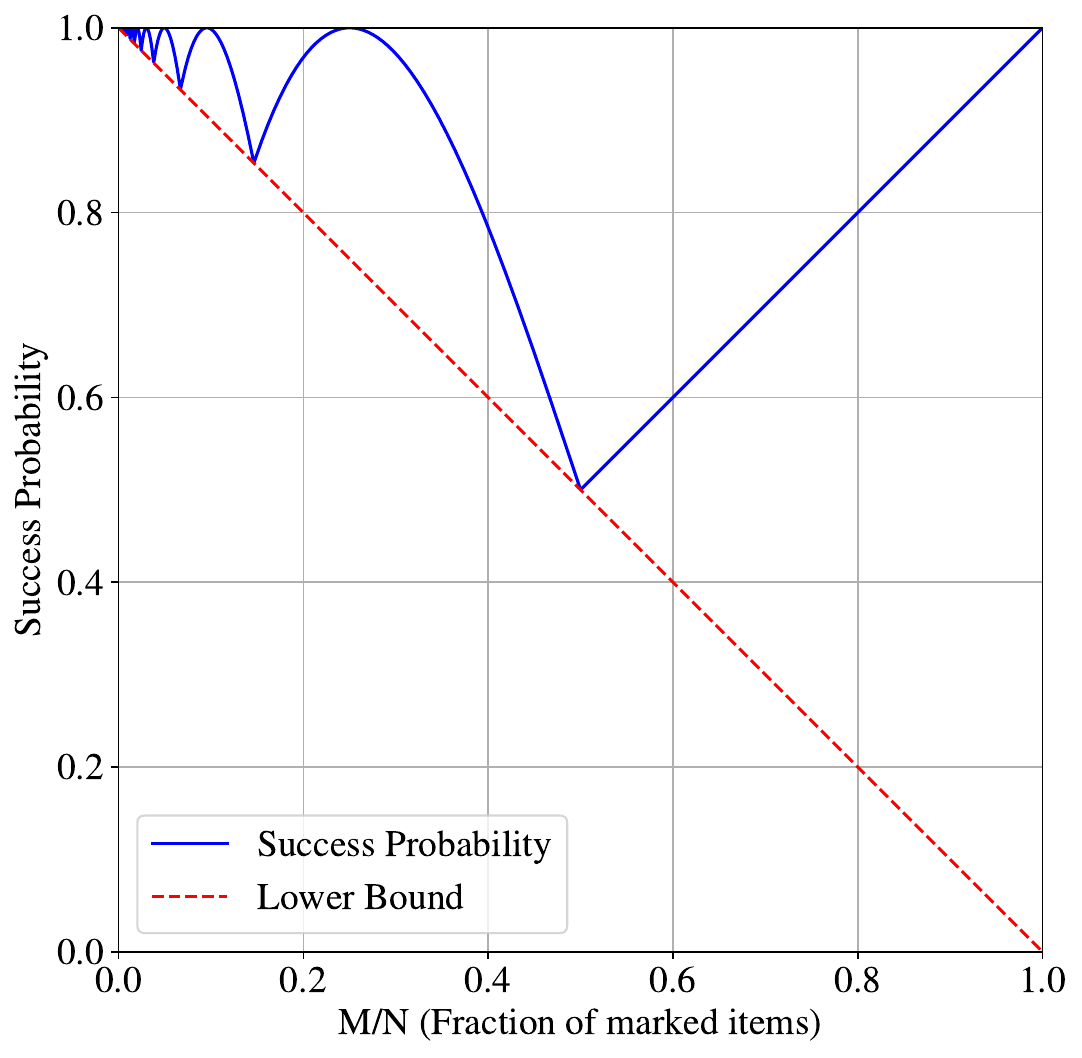}  
    \caption{The success probability of Grover's algorithm} 
    \label{fig:grover_succ_prob}   
\end{figure}

Moreover, even in the idealized case where $\theta$ is known with high precision, the optimal number of Grover iterations $k_{\mathrm{opt}}$ must be an integer. As a result, the quantum state evolves in discrete steps and, in general, cannot be rotated exactly to the optimal measurement point. Consequently, the final quantum state can only approach, but never exactly coincide with, the target state, as illustrated in Fig.~\ref{fig:grover_succ_prob}.

Taken together, the intrinsic uncertainty in quantum counting and the discreteness of the Grover iteration number imply that the success probability of Grover’s algorithm is generally below unity. This limitation persists even under ideal conditions and becomes more pronounced in realistic scenarios, where noise and model uncertainties further degrade the accuracy of the rotation-angle estimation.

Building upon the Grover quantum search algorithm, Long proposed an exact quantum search algorithm, known as the Long algorithm~\cite{long2001grover}. Under the assumption that the number of target items \( r \) in the database is known, this algorithm guarantees finding a target state with unit success probability. The core idea of the Long algorithm is to introduce phase matching to modify the Grover operator, such that the quantum state can be rotated exactly onto the target state within a finite number of iterations.

The Long operator is defined as
\begin{equation}\label{equ:long_ope}
L = - H^{\otimes n} R_0 H^{\otimes n} R_\tau ,
\end{equation}
where \( H^{\otimes n} \) denotes the \( n \)-qubit Hadamard transform.

The phase rotation operators generalize the \( D \) and \( U_f \) operators in the Grover algorithm as
\begin{equation}
R_0 = I + \left(e^{i\phi}-1\right)\ket{0}\bra{0}, \quad
R_\tau = I + \left(e^{i\psi}-1\right)\ket{\tau}\bra{\tau},
\end{equation}
where \( \ket{\tau} \) represents the target state. This modification replaces the \( \pi \)-phase inversion of the target state in the Grover oracle \( U_f \) with a general phase rotation \( \phi \), and generalizes the inversion-about-the-mean operation in \( U_s \) to a phase rotation \( \psi \) about the mean direction.

To ensure the validity of the algorithm, the Long algorithm requires the phase parameters \( \phi \) and \( \psi \) to satisfy the phase-matching condition \( \phi = \psi \). Under this condition, the rotation phase \( \phi \) can be determined as
\begin{equation}\label{phi_theta}
\phi = 2 \arcsin \left[ \frac{\sin\!\left( \frac{\pi}{4J_s+6} \right)}{\sin \theta}  \right],
\end{equation}
where \( \theta = \arcsin \sqrt{r/N} \), consistent with its definition in the Grover algorithm, \( N \) is the database size, and \( r \) is the number of target items. To ensure that the algorithm is well-defined, \( \phi \) must be real, which imposes the following constraint on \( J_s \):
\begin{equation}\label{equ:LONG_Js_min}
J_s > \left\lfloor \frac{\pi}{4\theta}-\frac{1}{2} \right\rfloor .
\end{equation}
It can be seen that the number of iterations in the Long algorithm is no less than that in the Grover algorithm. When \( J_s \) is less than or equal to this lower bound, the rotation phase \( \phi \) becomes complex, rendering the algorithm invalid.

Under the above condition, applying the Long operator \( L \) exactly \( J_s+1 \) times deterministically rotates the quantum state into the target subspace, such that measurement yields a target item with unit probability. The phase-matching condition is the key feature of the Long algorithm. In contrast, the fixed-phase setting in the Grover algorithm restricts the rotation of the quantum state, preventing it from reaching the target state exactly in a finite integer number of steps, and thus the success probability is generally less than 1. The Long algorithm overcomes this limitation by introducing adjustable phases to achieve exact phase matching.

The Long algorithm preserves the same asymptotic complexity as the Grover algorithm. The number of calls to the Long operator is \( J_s+1 \). In the optimal case, taking the lower bound of \( J_s \) that satisfies Eq.~\eqref{equ:LONG_Js_min}, we have
\begin{equation}\label{js_min}
    J_s = \left\lceil\frac{\pi}{4\theta}-\frac{1}{2}\right\rceil = \left\lceil\frac{\pi}{4\arcsin{\sqrt{r/N}}}-\frac{1}{2} \right\rceil .
\end{equation}

As in the Grover algorithm, for most search problems the number of target items is much smaller than the database size, i.e., \( r \ll N \). In this regime, \( \arcsin{\sqrt{r/N}} \approx \sqrt{r/N} \), yielding
\begin{equation}
    J_s = \frac{\pi}{4}\sqrt{\frac{N}{r}}-\frac{1}{2}=O\!\left(\sqrt{\frac{N}{r}}\right).
\end{equation}
Thus, the Long algorithm achieves the same asymptotic complexity as the Grover algorithm, while guaranteeing unit success probability and still providing a quadratic speedup over classical search.

From an implementation perspective, the quantum gate complexity of the Long algorithm should also be considered. Since each iteration only introduces a constant number of additional phase rotation gates, the gate complexity of the Long algorithm increases only by a constant factor compared to the Grover algorithm, and its asymptotic complexity remains unchanged.

Building upon Grover’s quantum search algorithm, Long proposed an exact quantum search algorithm, known as the Long algorithm~\cite{long2001grover}, which guarantees unit success probability for identifying the target state. The central idea of the Long algorithm is to modify the Grover operator by introducing phase rotations, such that the quantum state is rotated exactly onto the target subspace after a finite number of iterations.

The Long algorithm extends the two-dimensional SU(2) rotation of the Grover algorithm to a three-dimensional SO(3) rotation~\cite{long2001grover, zhao2012geometric}. The Bloch vector corresponding to the initial state of the system is given by
\begin{equation}
\vec r_i =
\begin{pmatrix}
\sin(2\theta) \\[1ex]
0 \\[1ex]
-\cos(2\theta)
\end{pmatrix},
\end{equation}
while the target state is
\begin{equation}
\vec r_f =
\begin{pmatrix}
0 \\[1ex]
0 \\[1ex]
1
\end{pmatrix}.
\end{equation}

The operator $L$ acts as a three-dimensional rotation matrix. Its action on the quantum state corresponds to a rotation about the axis
\begin{equation}
\vec l =
\begin{pmatrix}
\cos \frac{\phi}{2} \\[1ex]
\sin \frac{\phi}{2} \\[1ex]
\cos \frac{\phi}{2} \tan \theta
\end{pmatrix},
\end{equation}
with a rotation angle
\begin{equation}
\alpha = 4 \arcsin \big(\sin \frac{\phi}{2} , \sin \theta\big)=\frac{2\pi}{2J_s+3}.
\end{equation}
The total number of iterations is $k = J_s+1$, and the corresponding total rotation angle is
\begin{equation}
\gamma = (J_s+1)\alpha.
\end{equation}

When applying the Long algorithm to matched filtering, its execution still requires prior knowledge of the number of target items \( r \) in the database. To address this issue, a strategy similar to that used in the Grover algorithm can be adopted: quantum counting is employed to estimate the parameter \( \theta \), from which \( J_s \), \( \phi \), and \( \psi \) can be determined. The Long algorithm is then executed for the search. However, due to the statistical error inherent in quantum counting, the quantum matched filtering algorithm based on the Long algorithm cannot guarantee success in a single run. In the following, the error behaviors of quantum matched filtering schemes based on the Grover algorithm and the Long algorithm are analyzed and compared.

Consider a database of size \( N \) containing \( r \) target items. After applying the Grover operator \( k \) times, the quantum state evolves as
\begin{align}
G^k\ket{s}=\sin((2k+1)\theta)\ket{\beta}+\cos((2k+1)\theta)\ket{w_{\alpha}}.
\end{align}
A measurement at this stage yields a target item with probability
\begin{equation}
P_G = \sin^2\!\big((2k+1)\theta\big).
\end{equation}
When the number of iterations satisfies the optimal condition
\begin{equation}
(2k_{\mathrm{opt}}+1)\theta=\frac{\pi}{2},
\end{equation}
the success probability reaches its maximum. If there exists an estimation error in the parameter, such that the true angle is \( \theta+\delta\theta \), then the success probability becomes
\begin{equation}
\begin{aligned}
P_G(\theta + \delta\theta) 
&= \sin^2\big((2k+1)\theta + (2k+1)\delta\theta\big) \\
&= \sin^2\big((2k+1)\theta\big) \\ 
   &+ 2(2k+1) \sin((2k+1)\theta) \cos((2k+1)\theta)\, \delta\theta \\
&\quad + (2k+1)^2 \cos\big(2(2k+1)\theta\big)\, (\delta\theta)^2 + O((\delta\theta)^3).
\end{aligned}
\end{equation}

When the number of iterations is chosen close to optimal,
\begin{equation} 
\sin((2k+1)\theta) \approx 1, \quad \cos((2k+1)\theta) \approx 0,
\end{equation}
we obtain
\begin{align}
1 - P_G 
&\approx (2k+1)^2 (\delta\theta)^2 \cos\big(2(2k+1)\theta\big)
&\approx \frac{(2k+1)^2}{2} (\delta\theta)^2 \\
&= \frac{\pi^2}{8\theta^2}\delta \theta^2.
\end{align}
Therefore, in the limit \( \theta \ll 1 \), the success probability of the Grover algorithm is highly sensitive to errors in the estimation of \( \theta \), with the error amplified by a factor of \( 1/\theta^2 \).

According to the geometric interpretation of the Long algorithm, the rotated vector is given by
\begin{equation}
\vec r_f = \vec r_i \cos \gamma + (\vec l \times \vec r_i) \sin \gamma + \vec l (\vec l \cdot \vec r_i)(1 - \cos \gamma),
\end{equation}
and the final success probability is:~\cite{long2001grover}
\begin{equation}
P = \frac{1+z}{2},
\end{equation}
where \( z \) is the \( z \)-component of \( \vec r_f \).

The cross product and inner product are computed as
\begin{equation}
(\vec l \times \vec r_i)_z = - \sin \frac{\phi}{2} \, \sin(2\theta), 
\vec l \cdot \vec r_i = \cos \frac{\phi}{2} \, \tan \theta, 
l_z (\vec l \cdot \vec r_i) = \cos^2 \frac{\phi}{2} \, \tan^2 \theta.
\end{equation}

Thus, the final \( z \)-component is
\begin{equation}
z = - \cos(2\theta) \cos \gamma
    - \sin \frac{\phi}{2} \, \sin(2\theta) \sin \gamma
    + \cos^2 \frac{\phi}{2} \, \tan^2 \theta \, (1 - \cos \gamma).
\end{equation}

Introducing an initial angle error \( \delta \theta \),
\begin{equation}
\theta \to \theta + \delta \theta, \quad
z(\theta + \delta \theta) = z(\theta) + \frac{d z}{d\theta} \delta \theta + \frac{1}{2} \frac{d^2 z}{d\theta^2} (\delta \theta)^2 + \cdots,
\end{equation}
we note that under ideal conditions \( z(\theta) = 1 \) corresponds to an extremum, and thus the first-order term vanishes. The error is therefore dominated by the second-order term:
\begin{equation}
1-P_L \approx \frac{1}{2} \cdot \frac{d^2 z}{d\theta^2} (\delta \theta)^2.
\end{equation}

Expanding in the small-\( \theta \) regime, Since \( \delta\theta \) is also a small quantity, its effect on the integer-valued parameter \( J_s \) is negligible. Therefore, \( J_s \) can be regarded as a constant in the error analysis, the leading second-order error is obtained as
\begin{equation}
1 - P_L \approx \cos\!\Bigg(\frac{2 (1+J_s) \pi}{3 + 2 J_s}\Bigg) \, \delta \theta^2.
\end{equation}

This result shows that the Long algorithm exhibits strong robustness in the small-angle regime: the first-order error is completely canceled, and the second-order error coefficient is governed by \( \cos\!\big(2(1+J_s)\pi/(3+2J_s)\big) \), whose absolute value is at most 1. Therefore, the dominant error remains bounded and well-controlled. Compared with the standard Grover algorithm, the Long algorithm is less sensitive to angle estimation errors, demonstrating superior stability.

\textit{Long algorithm based quantum matched filtering\textemdash}
In the algorithm we proposed, the algorithm framework is the same as QMF in Ref.~\cite{gao2022quantum} but we replace the Grover's algorithm with Long algorithm and preserve the quantum counting part. For completeness, we briefly summarize the key elements of this framework below, for more details about the quantum matched filtering framework, please refer to Ref.~\cite{gao2022quantum}.

First we use quantum counting to check whether any matching signal exists, and second step applies Long algorithm to identify a matching template, when at least one match is present.

For convenience, the same notation as in Ref.~\cite{gao2022quantum} is used: the number of templates is denoted by $N$, and the number of data points in the time-series by $M$. They chose a digital encoding to represent the data and templates as classical bits encoded in the computational basis, making use of four registers: one data register $\ket{D}$ which must be of size linear in $M$, and one index register $\ket{I}$, which requires $\log N$ qubits. For intermediate calculations they specify also one register to hold the computed template $\ket{T}$, which must be of size linear in $M$, and one $\ket{\rho}$to hold the computed SNR value, which does not scale with $N$ or $M$ and is $O(1)$.

The oracle construction in Long's algorithm is similar to that in Grover's algorithm. The key modification is replacing the fixed phase reflections with tunable phase rotations in both the oracle and diffusion operators, enabling deterministic amplitude amplification when the number of marked states is known. Algorithm~\ref{alg:LongGate} constructs the oracle required for the Long algorithm. Similar to the Grover algorithm, the core of this oracle is still to apply a phase rotation to the templates that satisfy the matching condition. However, the original phase factor \( e^{i\pi} \) is replaced by a tunable phase \( e^{i\phi} \), and the diffusion operator is modified accordingly. Similar to the quantum matched filtering oracle in Ref.~\cite{gao2022quantum}, the computational procedure can be divided into several steps.

\begin{algorithm}[htbp] 
\caption{Long's Gate
\newline Complexity: $O(M\log M + \log N)$}
\label{alg:LongGate}
\begin{algorithmic}[1]

\Function{Long's Search Algorithm}{$N$, $\ket{D}$, $\rho_{\textrm{thr}}$, $r$}

\State \emph{Calculating the phase factor}:
\State $\theta \gets \arcsin \sqrt{r/N}, \quad J_s \gets \left\lceil\frac{\pi}{4\theta}-\frac{1}{2}\right\rceil, \quad \phi \gets 2 \arcsin \left[ \frac{\sin\!\left( \frac{\pi}{4J_s+6} \right)}{\sin \theta}  \right]$

\Procedure{Oracle Construction}{}
\State \emph{Creating templates}
\ForAll{$i < N$}
\State $\ket{i}\ket{0} \gets \ket{i}\ket{T_i}$
\EndFor

\State \emph{comparison with data:}
\State $\ket{i}\ket{D}\ket{T_i}\ket{0}
      \gets \ket{i}\ket{D}\ket{T_i}\ket{\rho(i)}$

\If{$\rho(i) < \rho_{\textrm{thr}}$}
\State $f(i)=0$
\Else
\State $f(i)=1$
\EndIf

\State $\ket{i}\ket{D}\ket{T_i}\ket{\rho(i)}
\gets e^{i\phi f(i)}\ket{i}\ket{D}\ket{T_i}\ket{\rho(i)}$

\State \emph{Dis-entangling Registers}
\State $e^{i\phi f(i)}\ket{i}\ket{D}\ket{T_i}\ket{\rho(i)}
      \gets e^{i\phi f(i)}\ket{i}\ket{D}\ket{T_i}\ket{0}$
\State $e^{i\phi f(i)}\ket{i}\ket{D}\ket{T_i}\ket{0}
      \gets e^{i\phi f(i)}\ket{i}\ket{D}\ket{0}\ket{0}$
\EndProcedure

\Procedure{Generalized Diffusion Operator}{}
\State $\sum e^{i\phi f(i)}\ket{i}
\gets \left[I-(1-e^{i\phi})\ket{s}\bra{s}\right]
      \sum e^{i\phi f(i)}\ket{i}$
\EndProcedure
\EndFunction
\end{algorithmic}
\end{algorithm}

\begin{algorithm}[htbp]
	\caption{Long's Template retrieval \newline Complexity: $O\left((M\log M + \log N)\cdot\sqrt{N}\right)$}
	\label{alg:templateretreivinglong}
	\begin{algorithmic}[1]
		\State $\textit{N} \gets \textrm{number of }\textit{templates}$
		\State $i \gets \textrm{index of} \textit{ templates}$
		\State $\rho_{\textrm{thr}} \gets$ \textit{ threshold}
		\State $\ket{0} \gets \textit{Data}$  $ \ket{D}$ 
		\State $r_{\ast}\gets\textrm{number of }\textit{matched templates}$
		\State \emph{Calculating the number of repetitions}:
		\State {$J_{s}\gets \textbf{Round}\left\lceil\frac{\pi}{4\arcsin{\sqrt{r_{\ast}/N}}}-\frac{1}{2} \right\rceil$}
		\Procedure{Retrieve one template}{}
		\Repeat
		\State Algorithm~\ref{alg:LongGate} \Call{Long's Gate}{$N$, $\ket{D}$, $\rho_{\textrm{thr}}$}, $(J_{s}+1)--$
		\Until{$(J_{s}+1)==0$}
		\State \emph{Output}:
		\State $i_{\textrm{correct}}$
		\EndProcedure
	\end{algorithmic}
\end{algorithm}

First, the quantum registers are initialized: the data are loaded and the index register is prepared in a uniform superposition state, with a cost of \( O(M+\log N) \). Then, the corresponding template waveform is generated according to the index in superposition, which requires \( O(M) \) operations. Next, the signal-to-noise ratio (SNR) is computed and compared with a given threshold, incurring a computational cost of \( O(M\log M) \). To ensure that the subsequent amplitude amplification operates correctly on the index register, the intermediate results, including the generated waveform and SNR, must be uncomputed (erased), which introduces an additional cost of \( O(M\log M) \). Finally, the Long algorithm applies a phase rotation to the states that satisfy the condition and performs the corresponding diffusion operation on the index register, with a cost of \( O(\log N) \).

Therefore, the total computational cost of a single Long oracle call is
\begin{equation}
\label{equ:longoracletotalcost}
O\left(M\log M + \log N\right).
\end{equation}

It can be seen that, compared with the Grover algorithm, although the Long algorithm modifies the phase rotation and amplitude amplification steps, the dominant computational cost of the oracle is still determined by template generation and SNR computation. Hence, its asymptotic complexity remains the same as that of the Grover algorithm.

Compared with the oracle of the Grover algorithm~\cite{gao2022quantum}, the additional overhead of the Long oracle lies in the gate complexity. Since each iteration only introduces a constant number of additional phase rotation gates, the gate complexity of the Long oracle increases only by a constant factor relative to the Grover case, without changing the asymptotic complexity.

First, quantum counting is employed to detect whether a matching signal exists. This step is identical to Algorithm in the original QMF~\cite{gao2022quantum}, and thus the complexity remains
\begin{equation}
	\label{longsignaldetectcost}
	O\left((M\log M + \log N)\cdot\sqrt{N}\right).
\end{equation}

Quantum counting provides an estimate \( r_\ast \) of \( r \), from which the phase factor \( \phi \) and the number of iterations \( J_s+1 \) of the Long algorithm can be determined using Eqs.~\eqref{phi_theta} and~\eqref{js_min}. Subsequently, Algorithm~\ref{alg:templateretreivinglong} employs the Long algorithm to search for the matching template, with total complexity
\begin{equation}
    O\left((M\log M + \log N)\cdot\sqrt{N}\right).
\end{equation}

Therefore, the overall complexity of the algorithm remains at the order of \( O(\sqrt{N}) \).

\begin{figure*}
\centering
\subfloat[Grover $\rho_{\textrm{thr}}=800$]{\label{figure_sim_g_800}\includegraphics[height=0.25\textwidth]{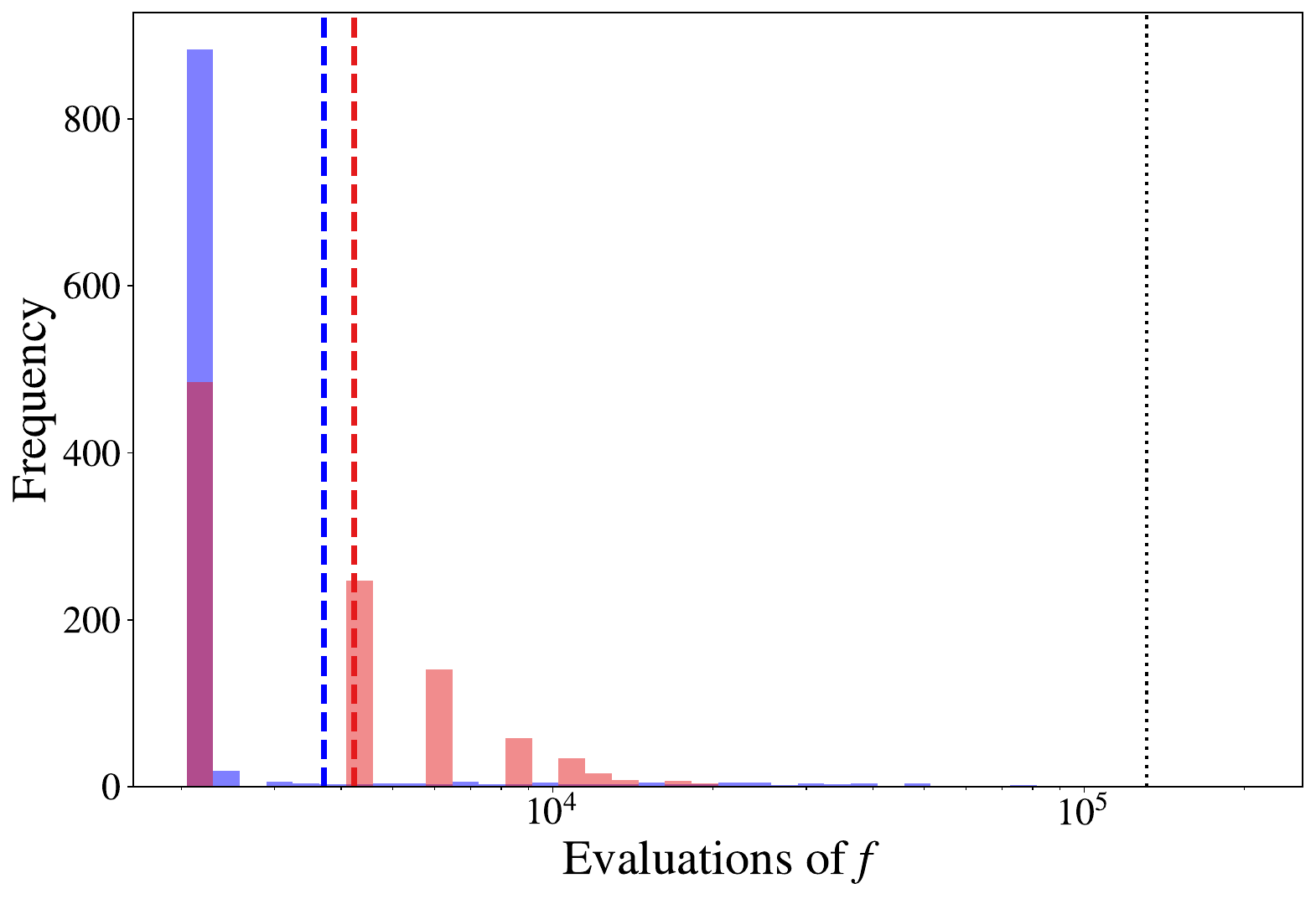}}
\subfloat[Grover $\rho_{\textrm{thr}}=850$]{\label{figure_sim_g_850}\includegraphics[height=0.25\textwidth]{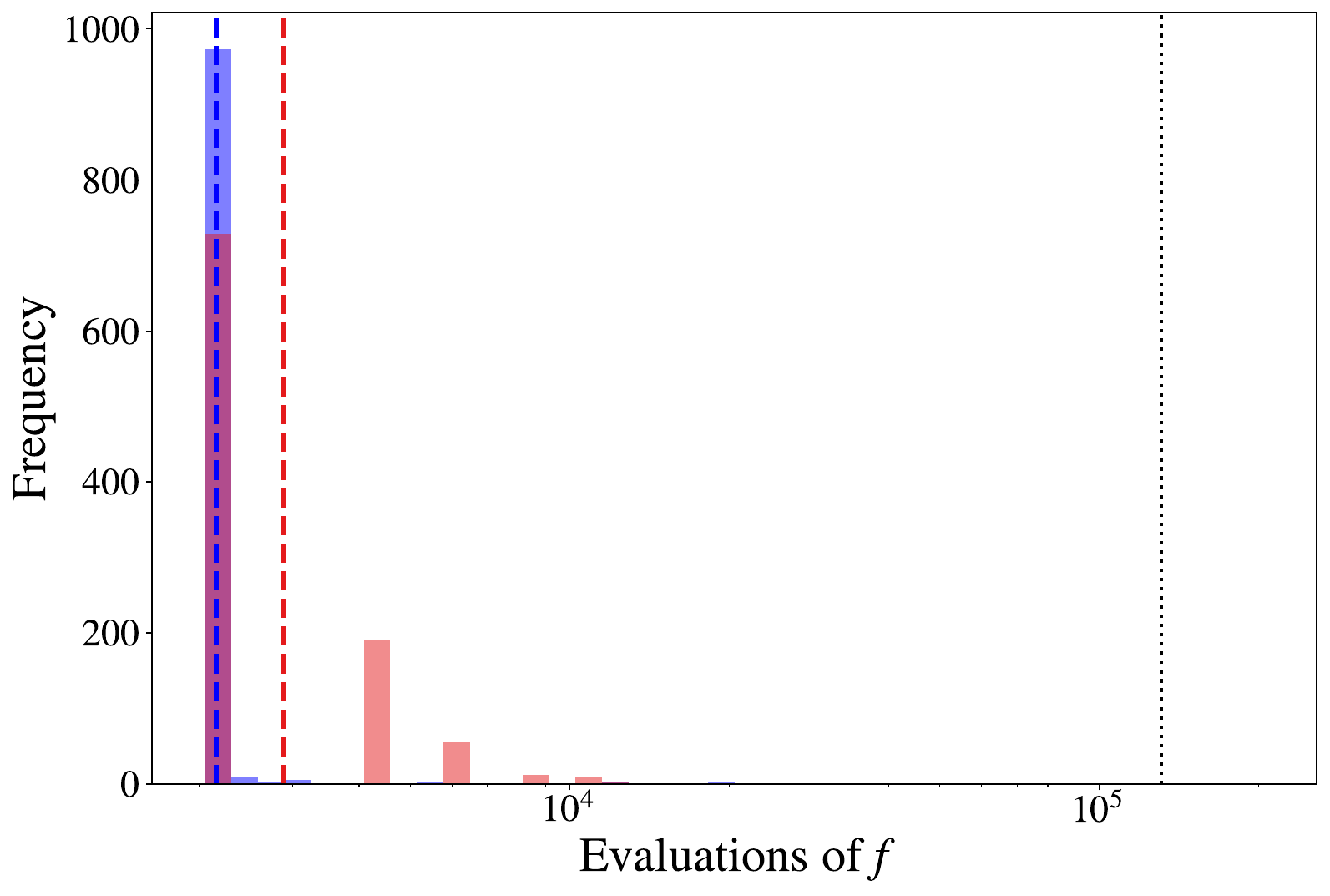}}\\

\subfloat[Long $\rho_{\textrm{thr}}=800$]{\label{figure_sim_l_800}\includegraphics[height=0.25\textwidth]{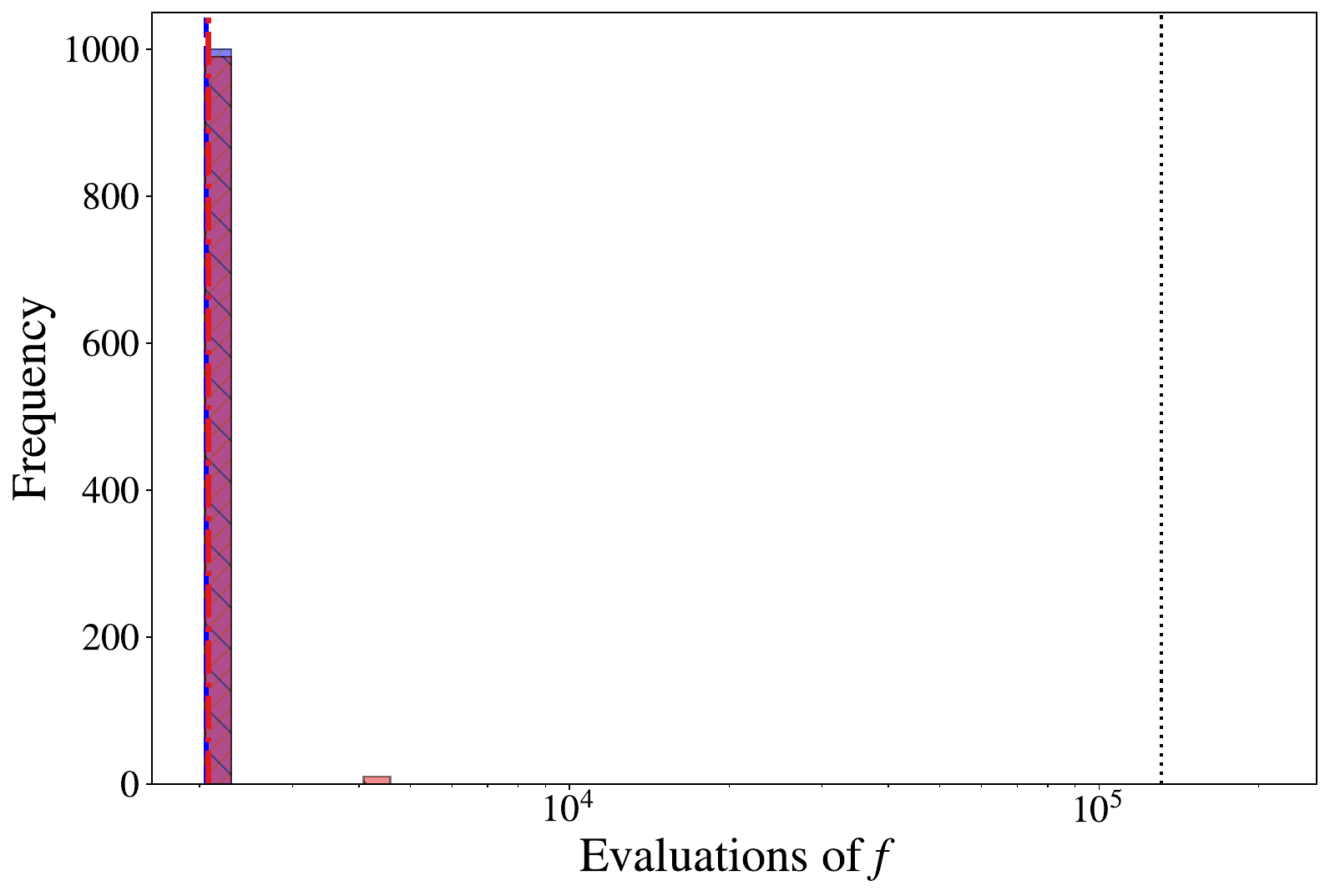}}
\subfloat[Long $\rho_{\textrm{thr}}=850$]{\label{figure_sim_l_850}\includegraphics[height=0.25\textwidth]{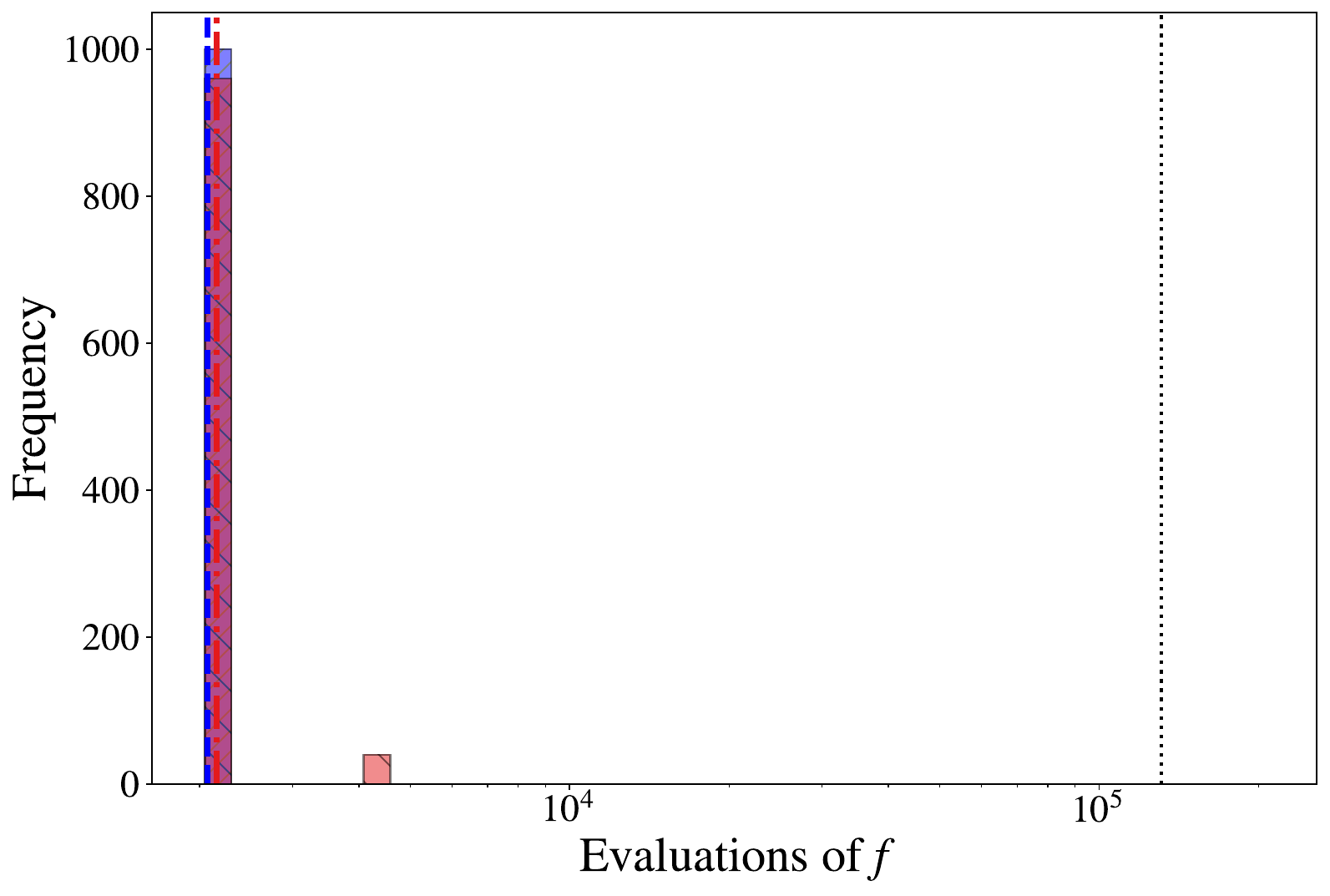}}\\

\subfloat[Grover $\rho_{\textrm{thr}}=860$]{\label{figure_sim_g_860}\includegraphics[height=0.25\textwidth]{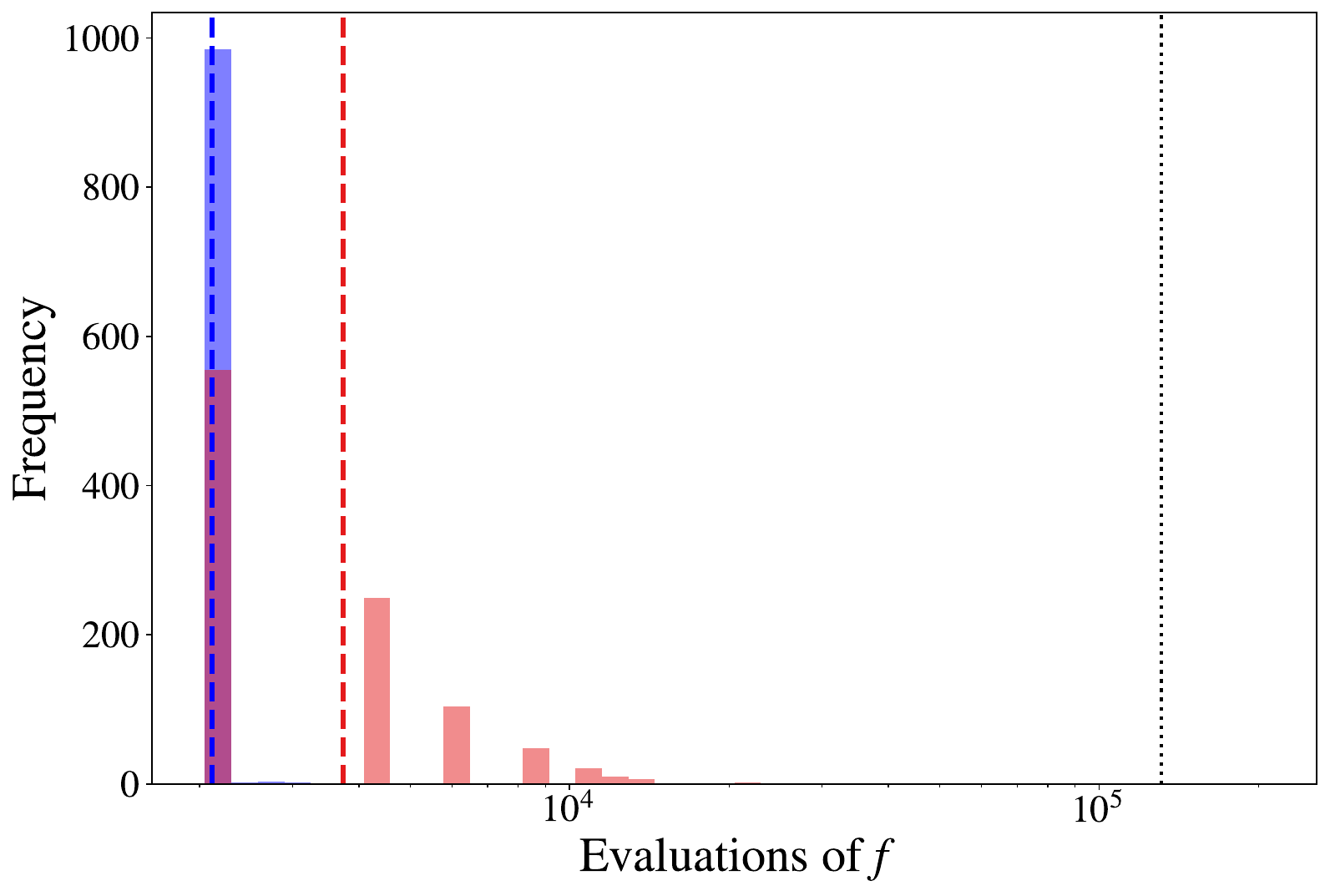}}
\subfloat[Grover $\rho_{\textrm{thr}}=870$]{\label{figure_sim_g_870}\includegraphics[height=0.25\textwidth]{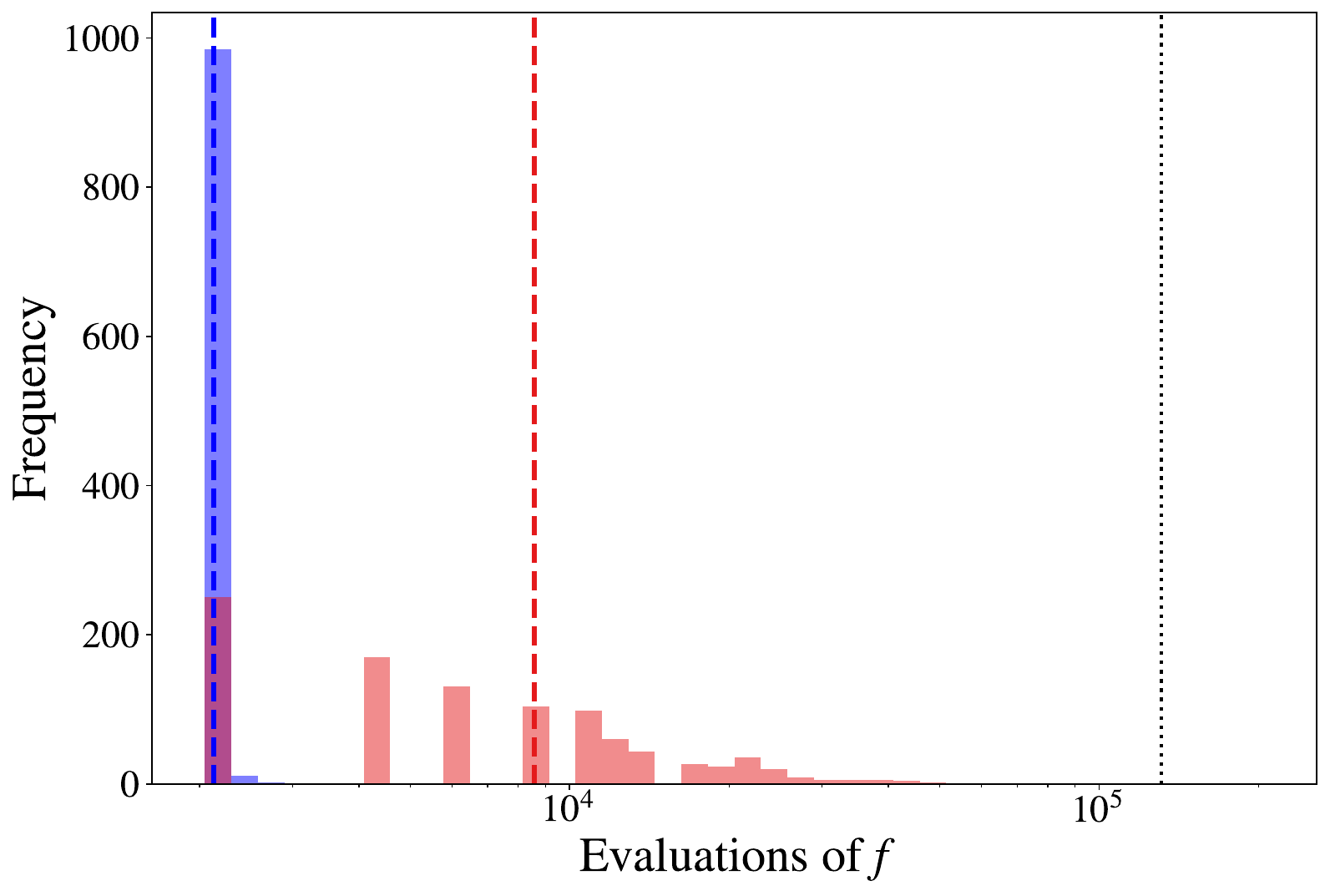}}\\

\subfloat[Long $\rho_{\textrm{thr}}=860$]{\label{figure_sima_l_860}\includegraphics[height=0.25\textwidth]{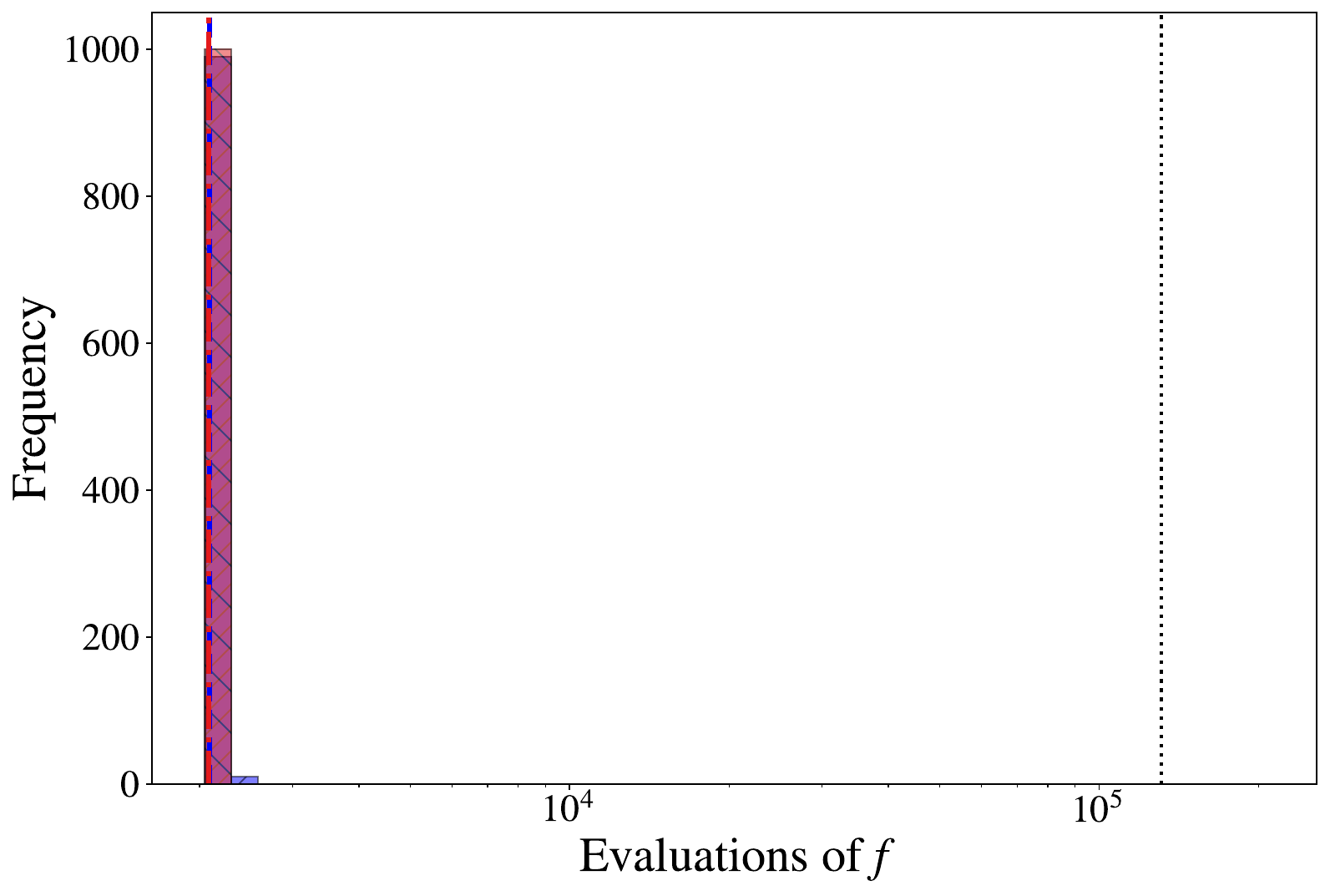}}
\subfloat[Long $\rho_{\textrm{thr}}=870$]{\label{figure_simb_l_870}\includegraphics[height=0.25\textwidth]{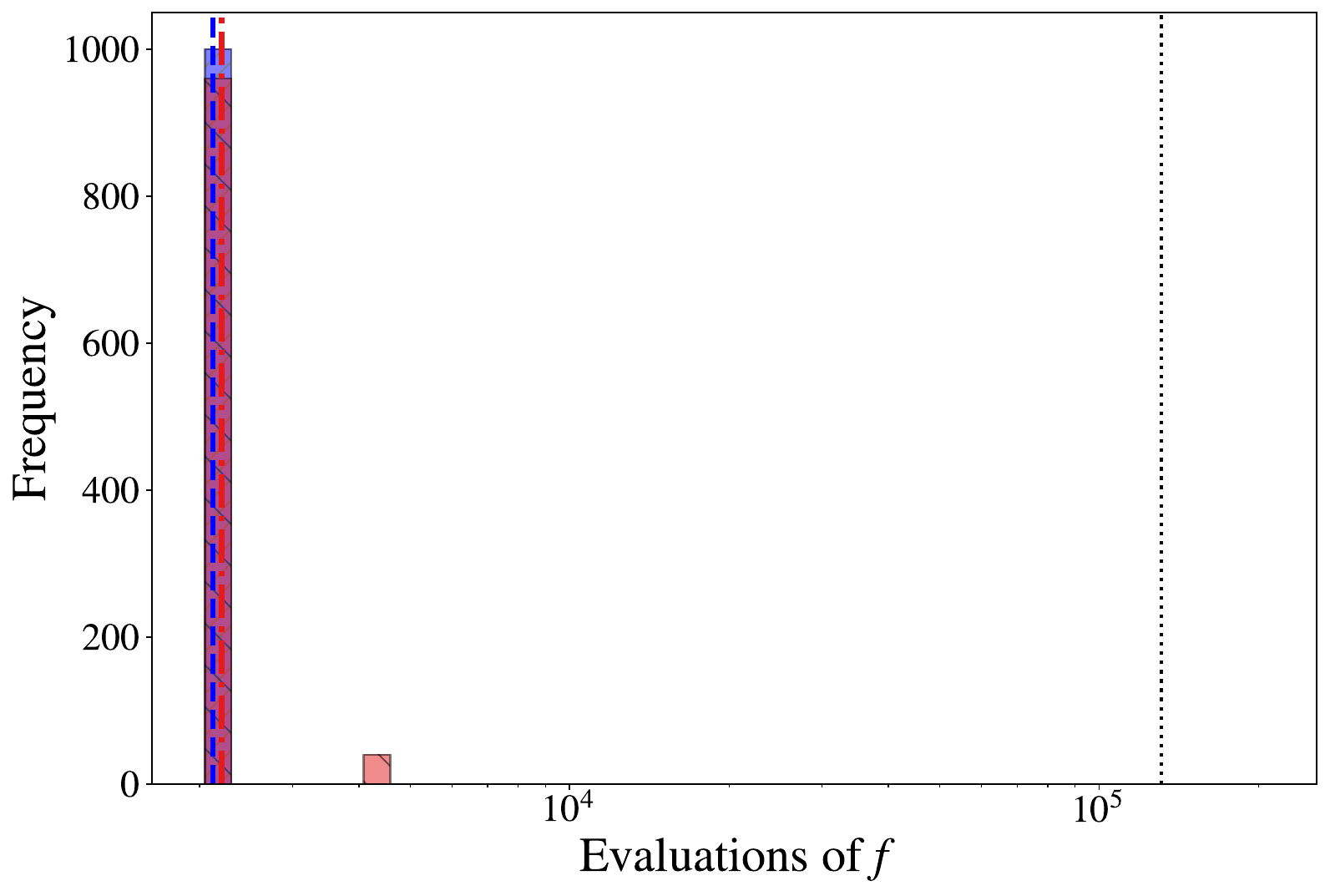}}

\caption{Comparison of the number of evaluations of the oracle function $f$ required to retrieve a matching template of Grover-based QMF and Long-based QMF in 1,000 simulations at  threshold  \( \rho_{\text{thr}} = 800, 850, 860, 870 \) for MBHB case. (a) (b) (e) (f) and (c) (d) (g) (h) subplot shows the distribution of function evaluations across trials of Grover-based QMF and
    Long-based QMF, respectively. Blue histograms represent using a fixed $ k_* $ estimated from a single \textsc{Signal Detection} step; red histograms correspond to re-estimating $ k_* $ for each failed retrieval. Dashed lines indicate mean values. The black dotted line shows the classical case where all $2^{17}$ templates are evaluated.}
\label{fig:sims_long_compar_thr}
\end{figure*}

\begin{figure*}
\centering
\subfloat[Grover $p=10$]{\label{figure_sim_10_800_g}\includegraphics[height=0.22\textwidth]{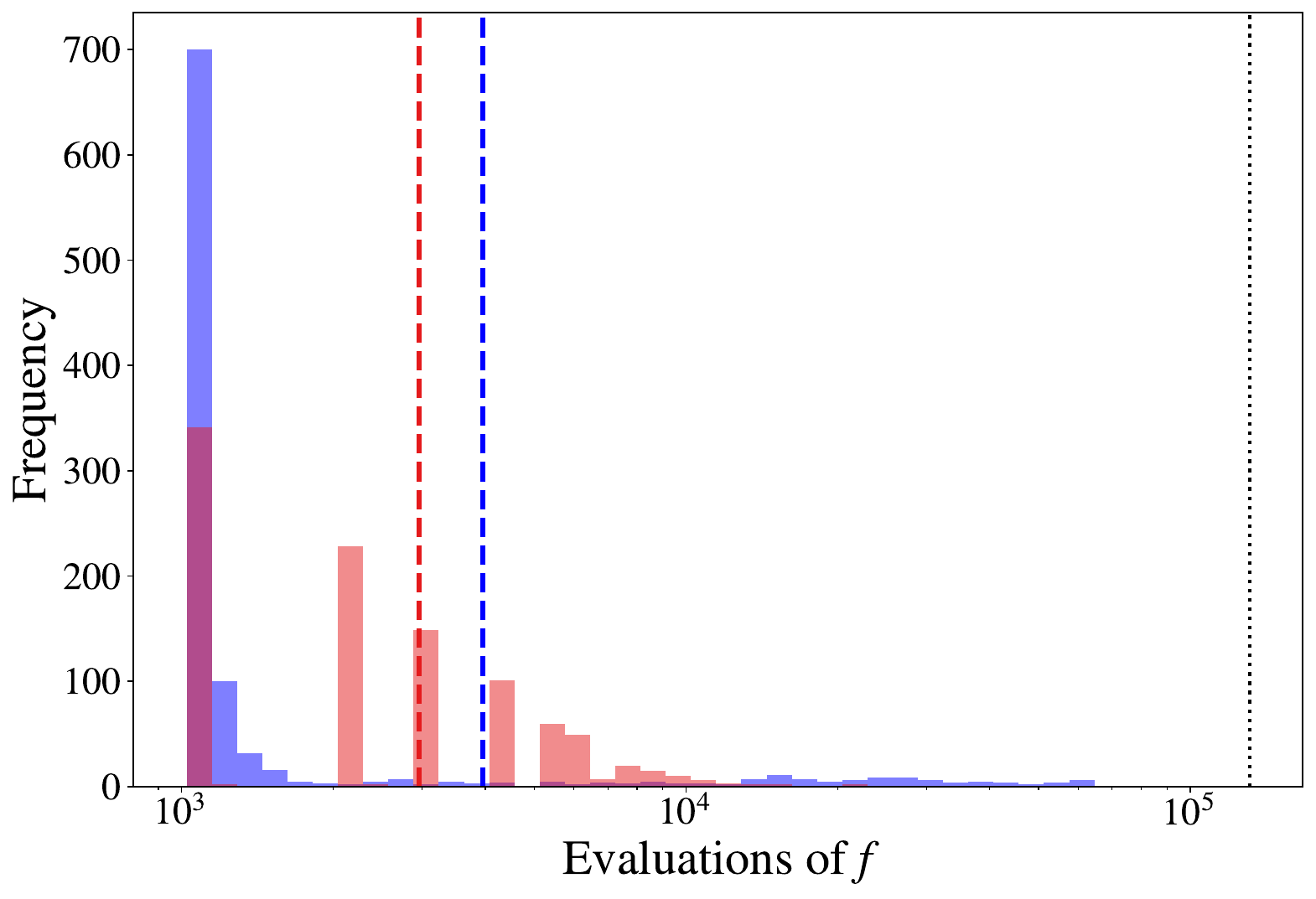}}
\subfloat[Grover $p=11$]{\label{figure_sim_11_800_g}\includegraphics[height=0.22\textwidth]{131072_2048_800_simulation_scenarios.pdf}}
\subfloat[Grover $p=12$]{\label{figure_sim_12_800_g}\includegraphics[height=0.22\textwidth]{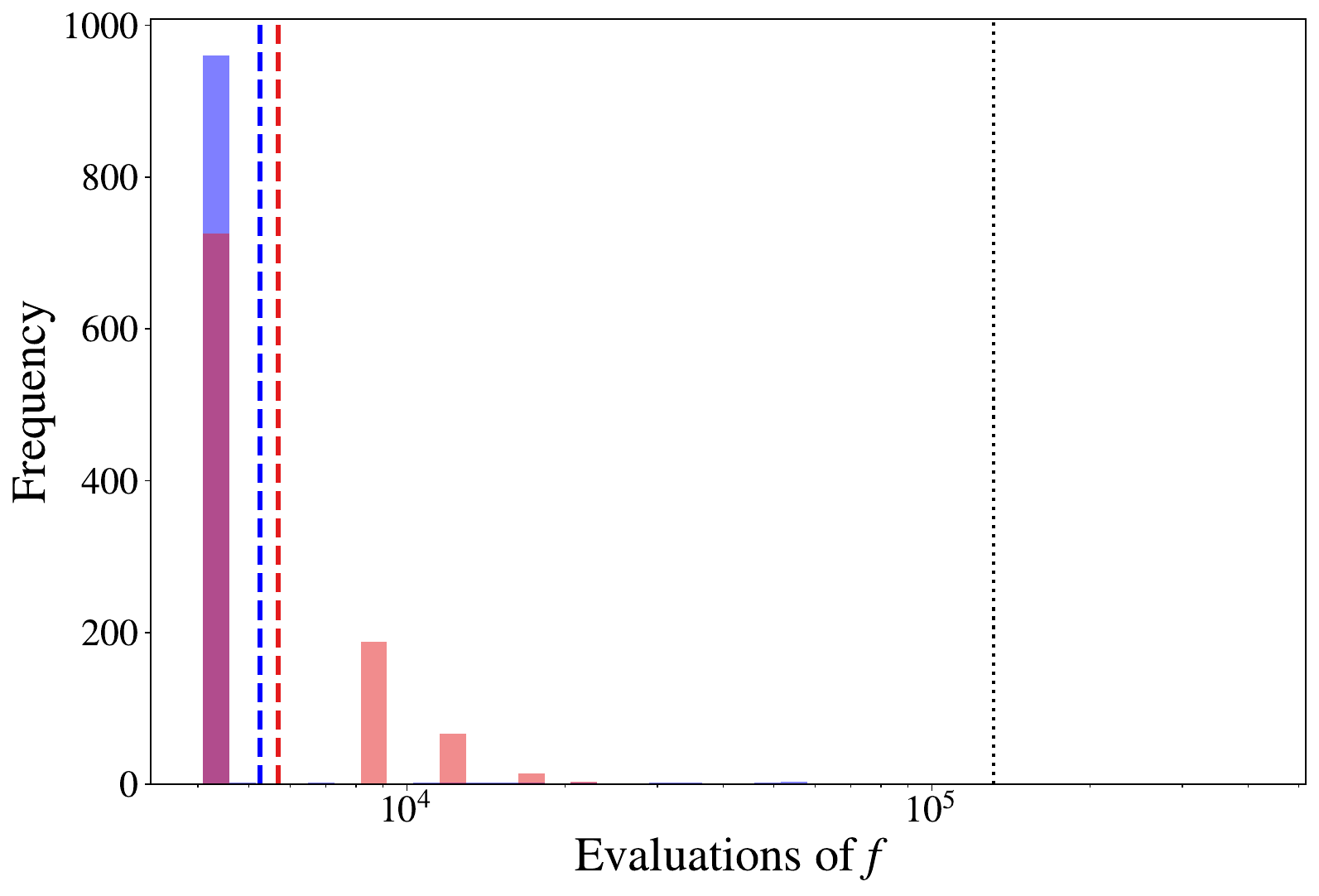}}\\
\subfloat[Long $p=10$]{\label{figure_sim_10_800_l}\includegraphics[height=0.22\textwidth]{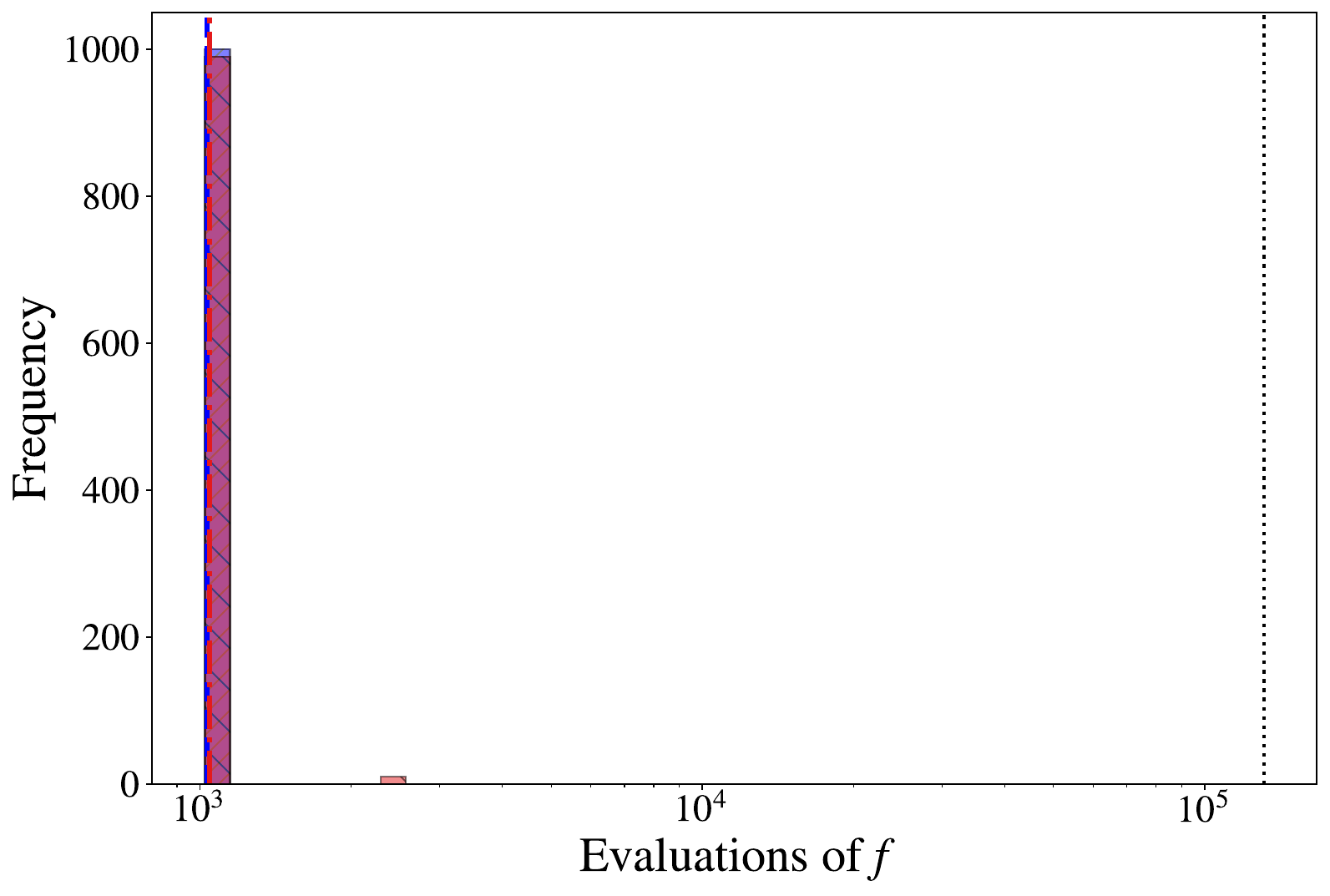}}
\subfloat[Long $p=11$]{\label{figure_sim_11_800_l}\includegraphics[height=0.22\textwidth]{2048_131072_800.0_simulation_long_counts_multi_two.pdf}}
\subfloat[Long $p=12$]{\label{figure_sim_12_800_l}\includegraphics[height=0.22\textwidth]{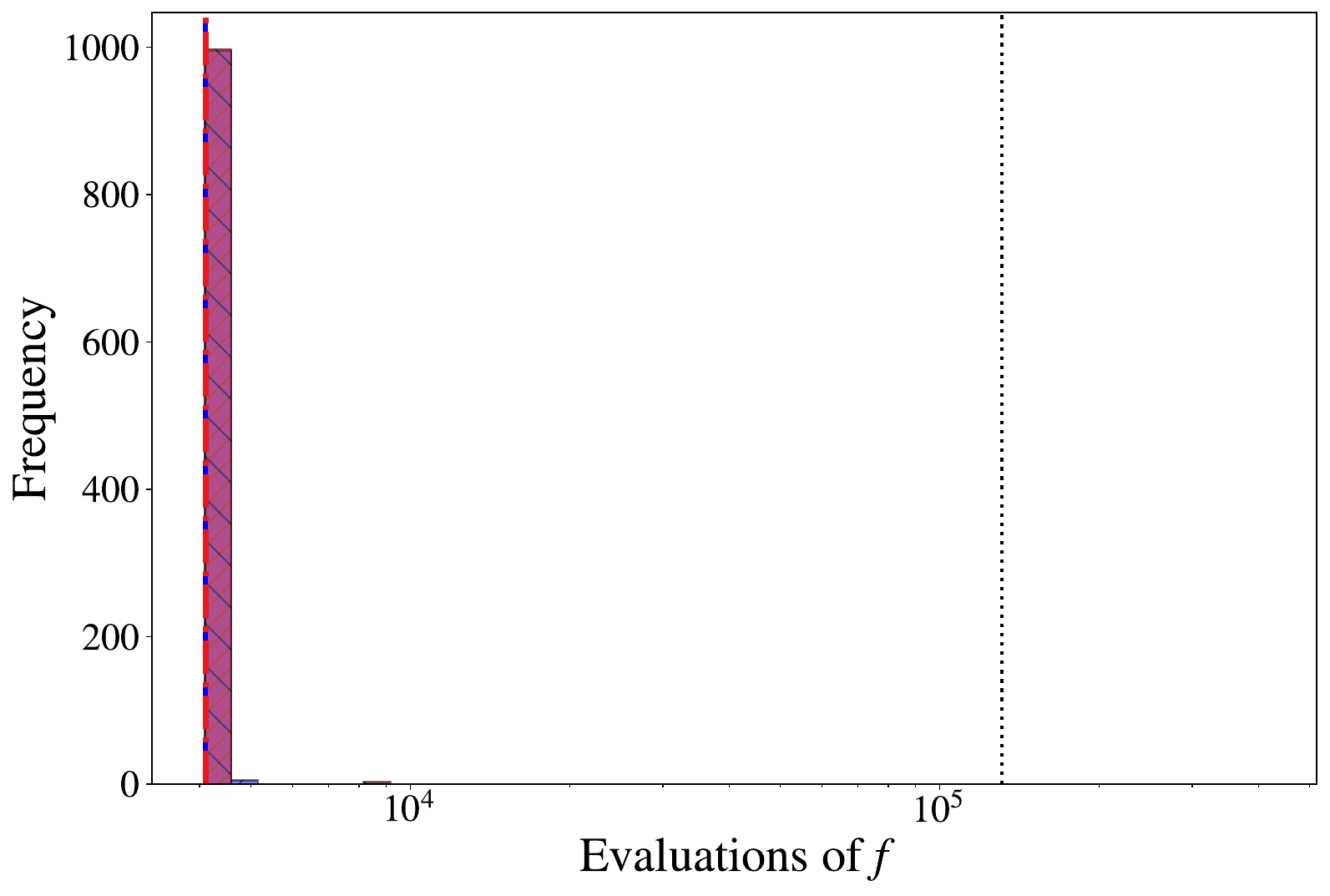}}
\caption{Comparison of the number of evaluations of the oracle function $f$ required to retrieve a matching template of Grover-based QMF and Long-based QMF in 1,000 simulations precision parameter $p=10,11,12$ for $\rho_{\text{thr}} = 800$  MBHB case. (a-c) and (d-f) subplot shows the distribution of function evaluations across trials of Grover-based QMF and
    Long-based QMF, respectively. Blue histograms represent using a fixed $ k_* $ estimated from a single \textsc{Signal Detection} step; red histograms correspond to re-estimating $ k_* $ for each failed retrieval. Dashed lines indicate mean values. The black dotted line shows the classical case where all $2^{17}$ templates are evaluated.
}
\label{fig:sims800_p_vary_g_l}
\end{figure*}

\textit{Simulation details and results\textemdash}
Since the quantum counting procedure is identical to that in the original algorithm, its output remains the same. Therefore, the differences between the two algorithms mainly arise in the subsequent template retrieval stage. Based on this, this part focuses only on the performance differences between the quantum matched filtering algorithms based on the Grover algorithm and the Long algorithm during the template retrieval stage. The test cases remain consistent with our privous work~\cite{guo2025quantum}, and the simulation still models the algorithm evolution by computing the probability amplitudes of quantum states, with the only modification being the replacement of the Grover search step by the Long search step.The simulation is built upon the \href{https://github.com/Fergus-Hayes/quantum-matched-filter}{\textit{quantum-matched-filter}} \textit{Python} code~\cite{pythoncode}, which accompanies Ref.~\cite{gao2022quantum}. The gravitational wave strain data used in this study is generated using BBHX package~\cite{katz2020gpu, katz2022fully, michael_katz_2021_5730688} and simulates the signal from a binary black hole system with component masses of $10^6\,M_\odot$ and $5 \times 10^5\,M_\odot$. The time series spans 3 days and is sampled at 0.1~Hz.

Two types of tests were conducted for Grover-based QMF in our previous work~\cite{guo2025quantum}: (i) evaluating the sensitivity of the algorithm to variations in the detection threshold \( \rho_{\textrm{thr}} \), and (ii) assessing the robustness of the algorithm to quantum counting errors by varying the precision of quantum counting. The results showed that the algorithm is sensitive to both factors. Therefore, under the same test cases and experimental setup, the template retrieval process based on the Long algorithm is evaluated to assess its robustness and overall performance under identical conditions.

The specific testing schemes are consistent with those in our previous work~\cite{guo2025quantum} used for the Grover algorithm.

Scheme (1): First, in the \textsc{Signal Detection} step, an estimate \( \theta_\ast \) of the target ratio angle is obtained. Then, the smallest integer \( J_s \) satisfying Eq.~\eqref{equ:LONG_Js_min} is selected to achieve high search efficiency, and the phase factor \( \phi \) is computed accordingly. Based on this, the \textsc{Template Retrieval} step of the Long algorithm is repeatedly executed until a matching template is found, while recording the number of oracle function \( f \) calls as a measure of algorithmic complexity.

Scheme (2): If a single Long search fails, the quantum counting procedure is repeated to obtain a new estimate of \( \theta \). The phase parameters are then recalculated, and the Long algorithm is executed again. This process is repeated until a matching template is successfully retrieved.

\begin{figure*}
\centering
\subfloat[Grover $p=10$]{\label{figure_sim_10_870_g}\includegraphics[height=0.21\textwidth]{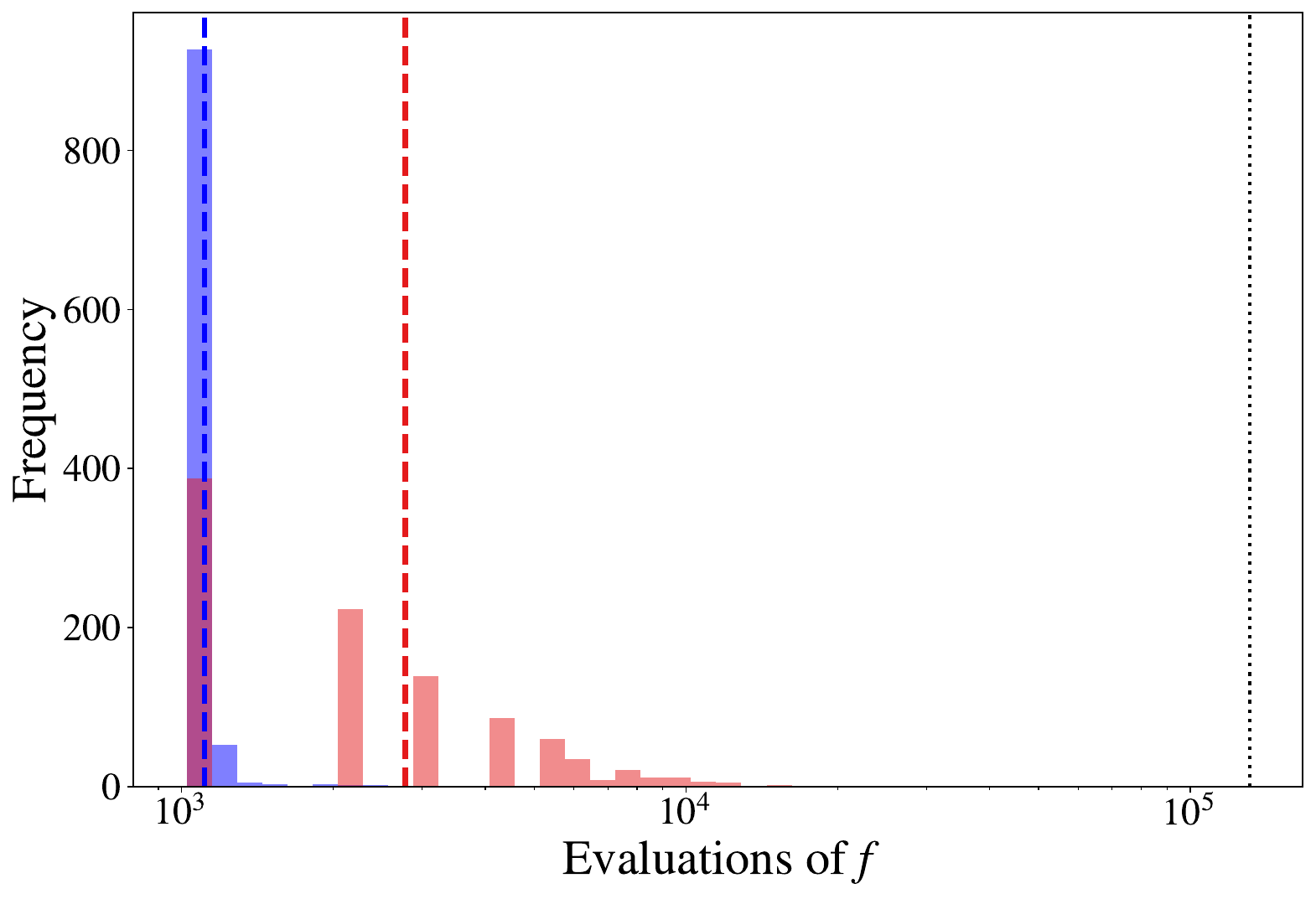}}
\subfloat[Grover $p=11$]{\label{figure_sim_11_870_g}\includegraphics[height=0.21\textwidth]{131072_2048_870_simulation_scenarios.pdf}}
\subfloat[Grover $p=12$]{\label{figure_sim_12_870_g}\includegraphics[height=0.21\textwidth]{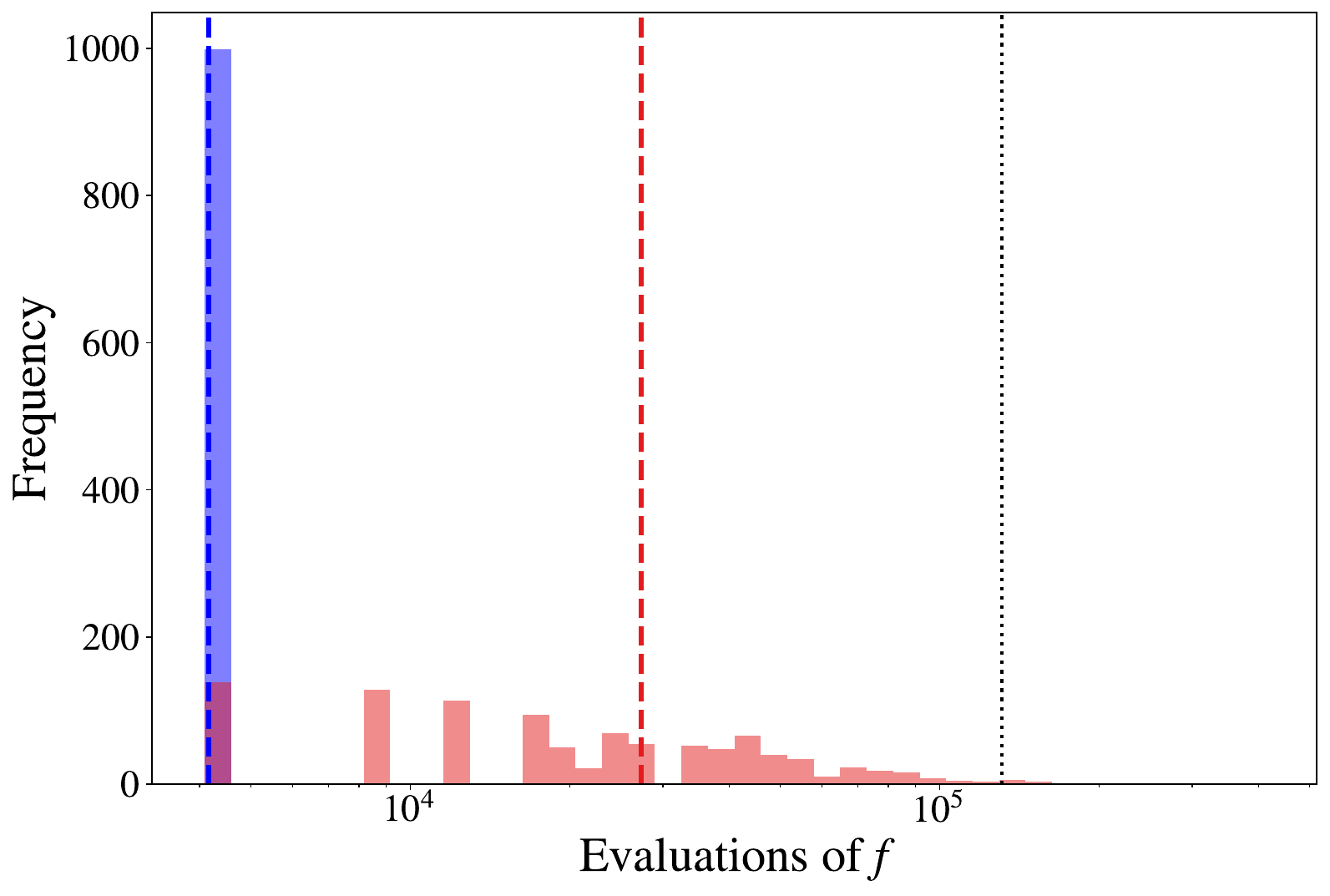}}\\
\subfloat[Long $p=10$]{\label{figure_sim_10_870_l}\includegraphics[height=0.21\textwidth]{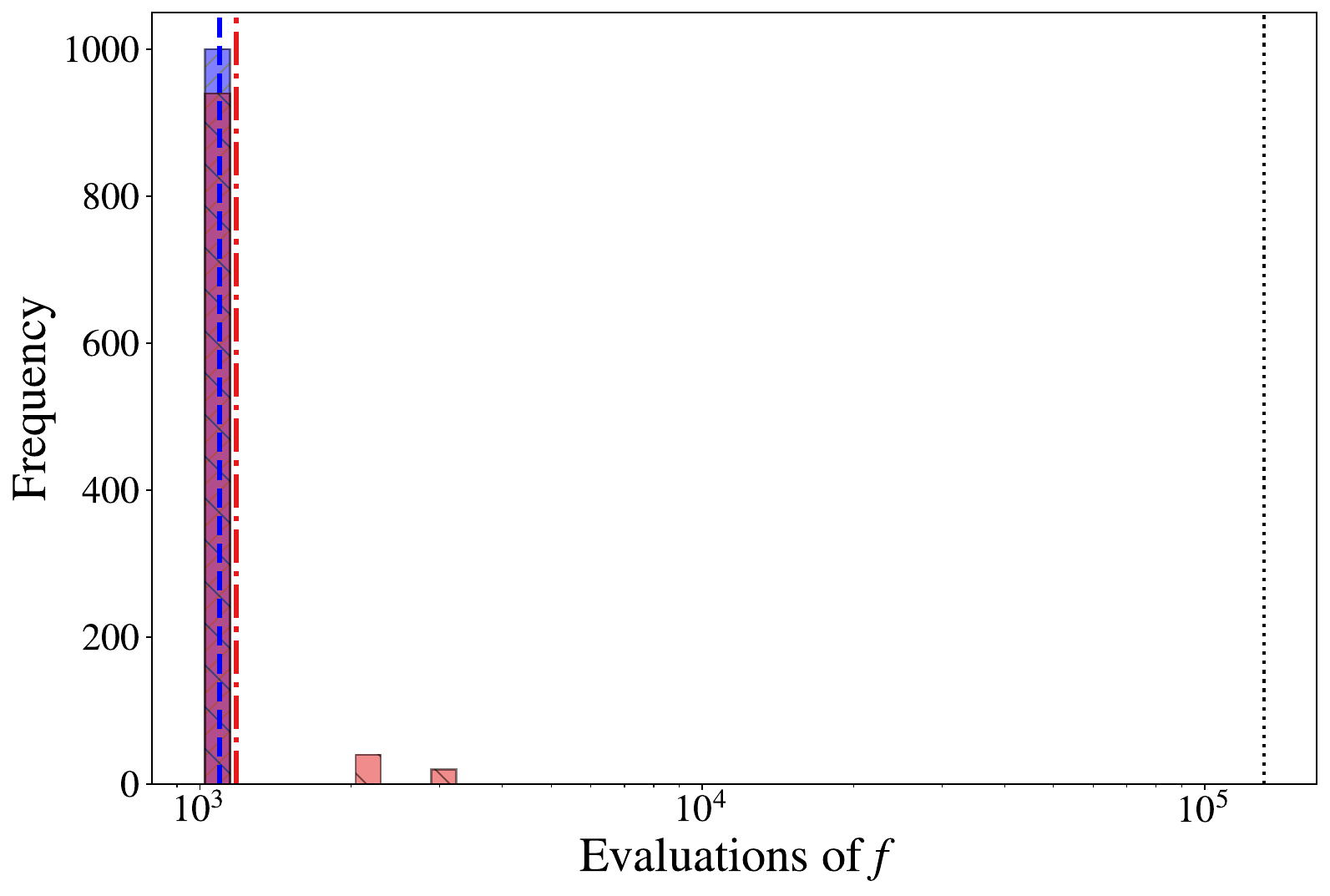}}
\subfloat[Long $p=11$]{\label{figure_sim_11_870_l}\includegraphics[height=0.21\textwidth]{2048_131072_870.0_simulation_long_counts_multi_two.pdf}}
\subfloat[Long $p=12$]{\label{figure_sim_12_870_l}\includegraphics[height=0.21\textwidth]{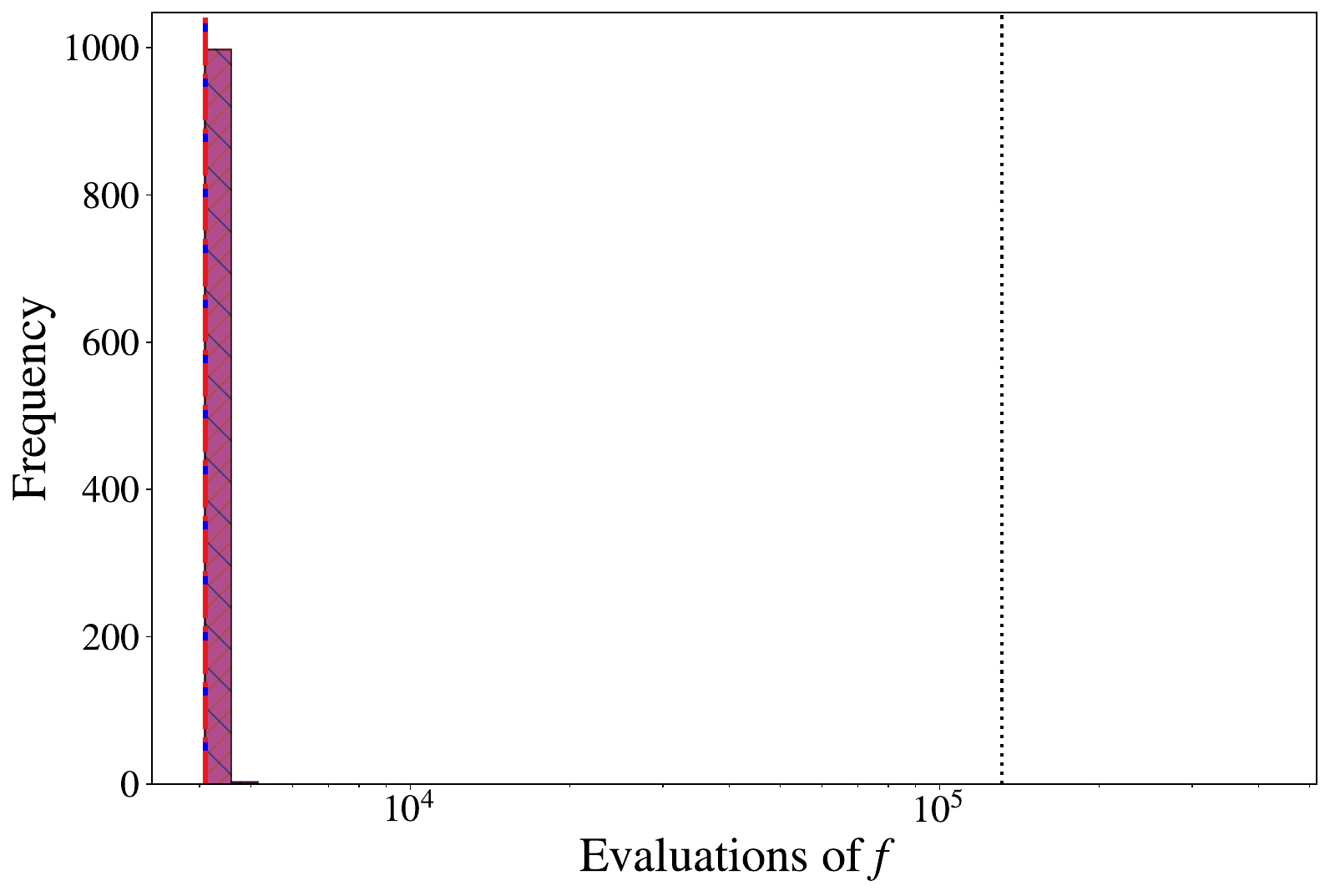}}
\caption{Comparison of the number of evaluations of the oracle function $f$ required to retrieve a matching template of Grover-based QMF and Long-based QMF in 1,000 simulations precision parameter $p=10,11,12$ for $\rho_{\text{thr}} = 870$  MBHB case. (a-c) and (d-f) subplot shows the distribution of function evaluations across trials of Grover-based QMF and
    Long-based QMF, respectively. Blue histograms represent using a fixed $ k_* $ estimated from a single \textsc{Signal Detection} step; red histograms correspond to re-estimating $ k_* $ for each failed retrieval. Dashed lines indicate mean values. The black dotted line shows the classical case where all $2^{17}$ templates are evaluated.}
\label{fig:sims870_p_vary_g_l}
\end{figure*}

\begin{figure*}
\centering
\subfloat[Grover $\rho_{\text{thr}}=18$]{\label{original_result-g}\includegraphics[height=0.25\textwidth]{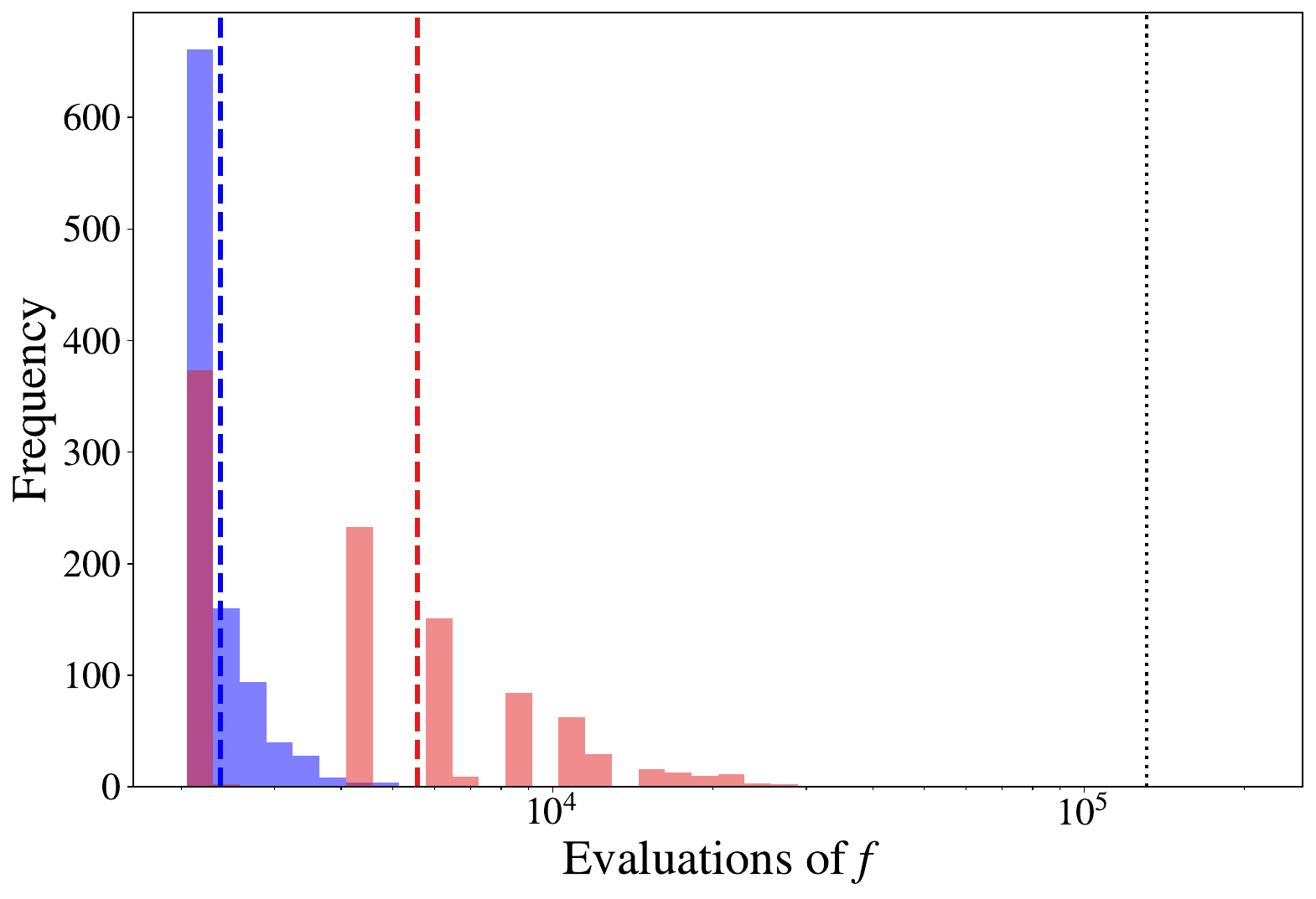}}
\subfloat[Grover $\rho_{\text{thr}}=16$]{\label{thr_16-g}\includegraphics[height=0.25\textwidth]{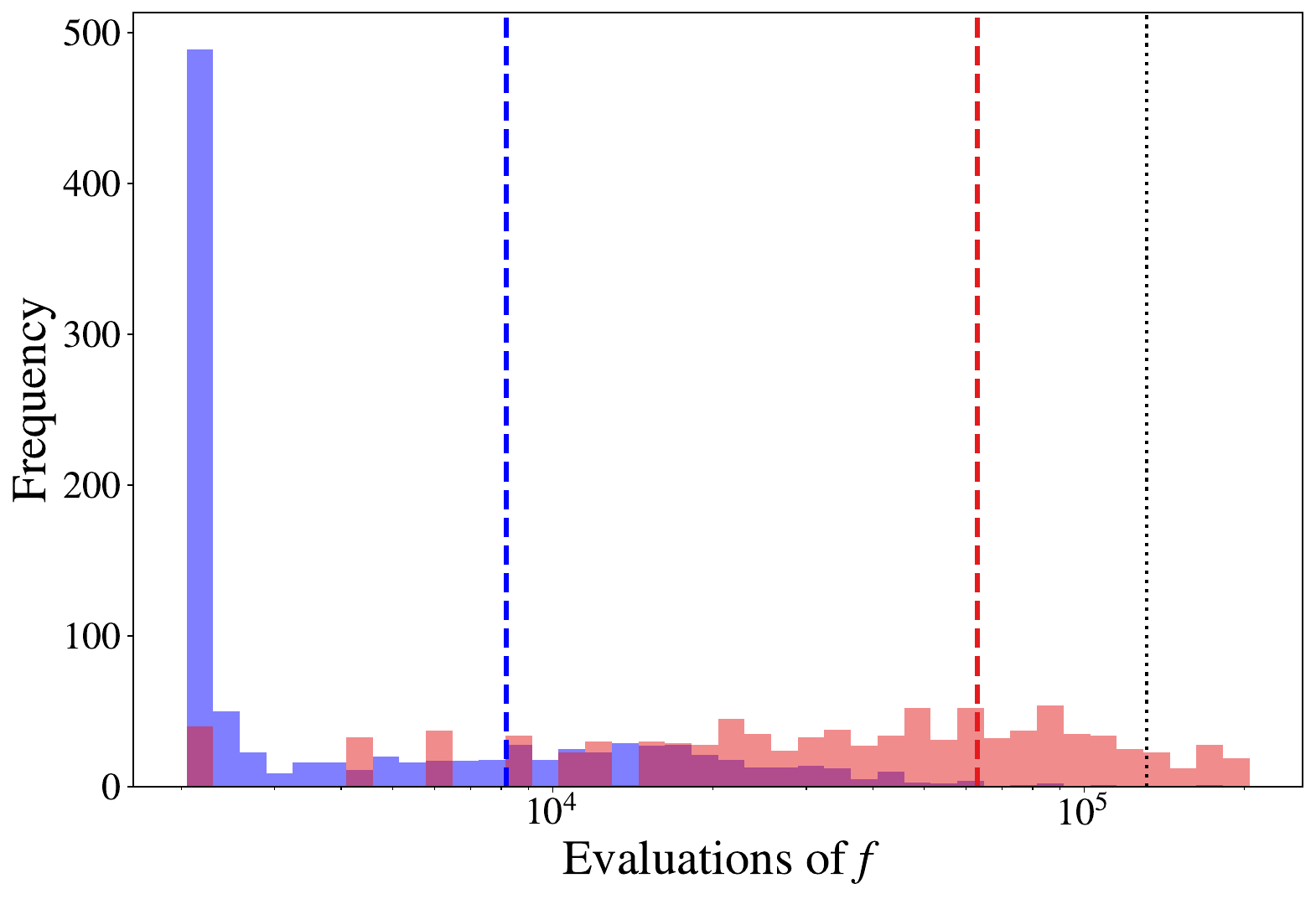}}\\
\subfloat[Long $\rho_{\text{thr}}=18$]{\label{original_result-l}\includegraphics[height=0.25\textwidth]{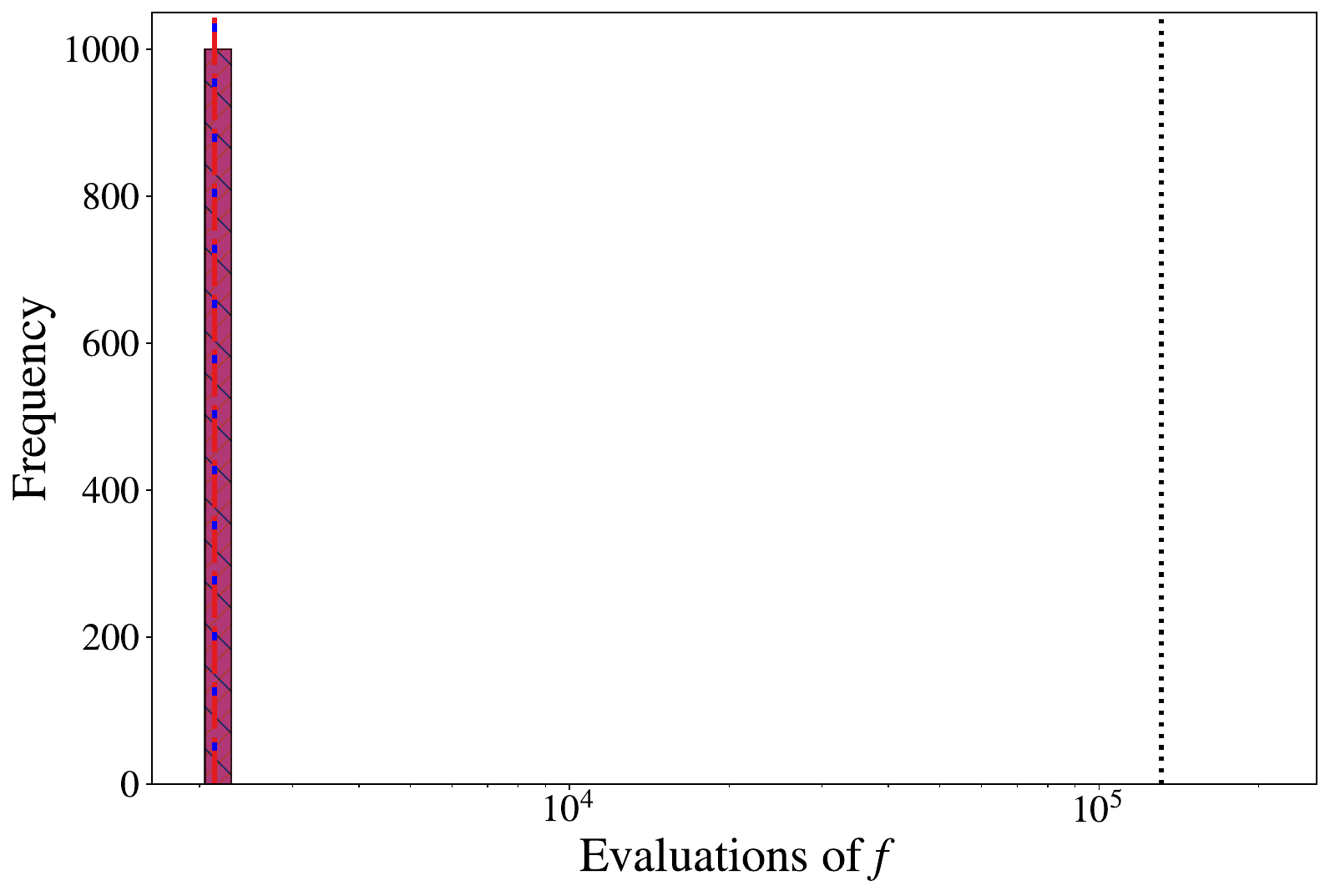}}
\subfloat[Long $\rho_{\text{thr}}=16$]{\label{thr_16-l}\includegraphics[height=0.25\textwidth]{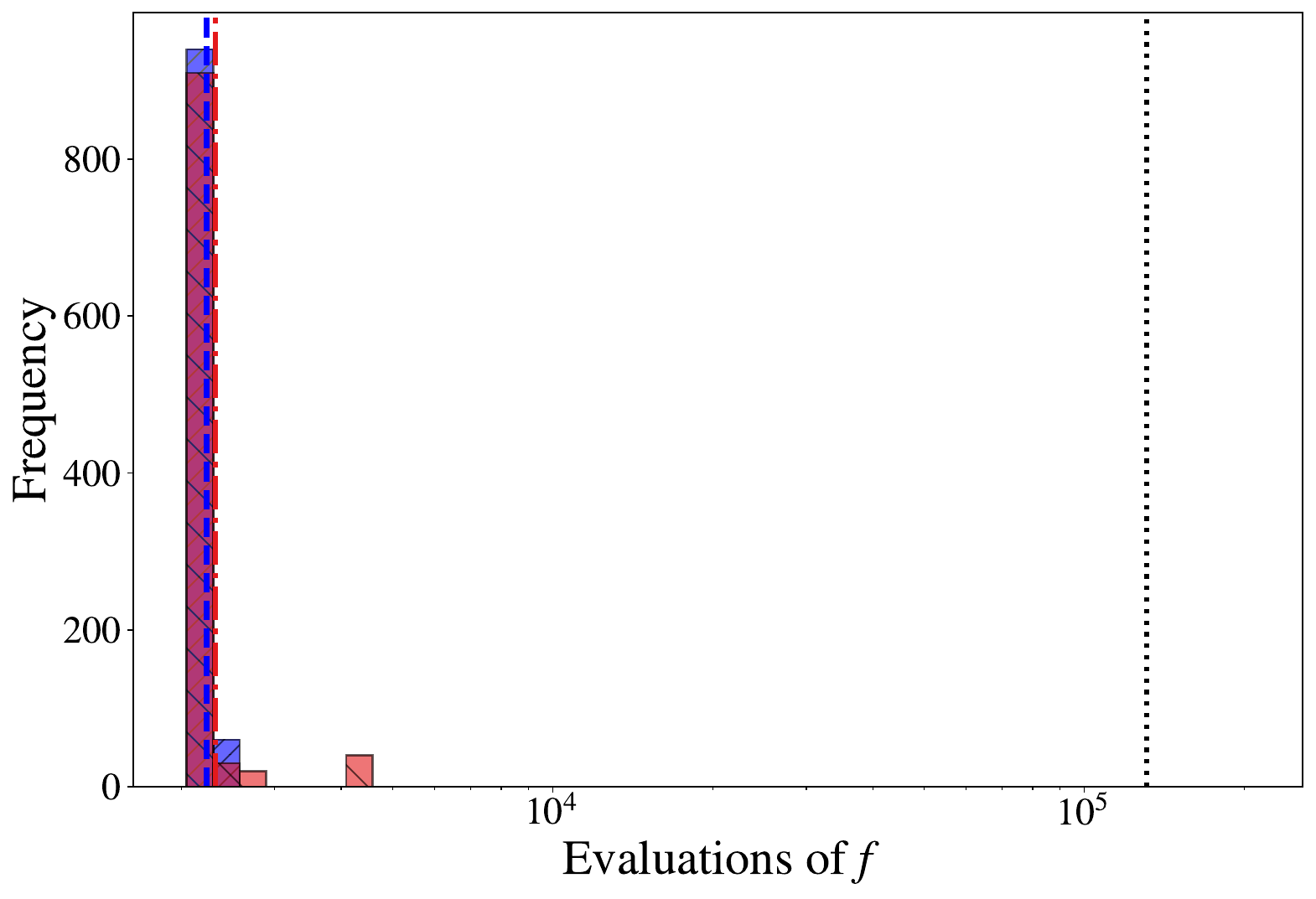}}
\caption{Comparison of the number of evaluations of the oracle function $f$ required to retrieve a matching template of Grover-based QMF and Long-based QMF in 1,000 simulations at  threshold  \( \rho_{\text{thr}} = 18, 16 \) for GW150914 case. (a) (b)  and (c) (d)  subplot shows the distribution of function evaluations across trials of Grover-based QMF and
    Long-based QMF, respectively. Blue histograms represent using a fixed $ k_* $ estimated from a single \textsc{Signal Detection} step; red histograms correspond to re-estimating $ k_* $ for each failed retrieval. Dashed lines indicate mean values. The black dotted line shows the classical case where all $2^{17}$ templates are evaluated.}
\label{fig:gw150914_g_l}
\end{figure*}

Figure~\ref{fig:sims_long_compar_thr} shows the distribution of oracle function calls over 1000 simulations for both Grover and Long algorithms under different threshold values. Subfigures (a), (b), (e), and (f) correspond to the Grover algorithm with thresholds \( \rho_{\textrm{thr}} = 800, 850, 860, 870 \), respectively, while the corresponding Long algorithm results are shown in subfigures (c), (d), (g), and (h). The blue histograms represent the results of Scheme (1), and the red histograms correspond to Scheme (2). The dashed blue and red lines indicate the mean values for the two schemes, while the black dashed line represents the computational cost of classical exhaustive search.

It can be observed that under different detection thresholds \( \rho_{\textrm{thr}} \), the two quantum search algorithms exhibit significantly different distributions of computational complexity during the template retrieval stage. Overall, the results of the Grover algorithm are more dispersed, whereas the Long algorithm shows a highly concentrated distribution of oracle calls, indicating better stability and robustness.

We first briefly review the Grover algorithm results. As shown in subfigures (a), (b), (e), and (f), under different threshold conditions, the distribution of oracle calls exhibits a pronounced right-skewed long-tail structure. While most simulation runs are concentrated in a lower range of oracle calls, a non-negligible fraction requires significantly more calls, forming a long tail extending to the right. This phenomenon reflects the sensitivity of the Grover algorithm to the threshold \( \rho_{\textrm{thr}} \).

In contrast, the Long algorithm exhibits markedly different behavior. As shown in subfigures (c), (d), (g), and (h), under all threshold conditions, the distribution of oracle calls is highly concentrated, with the vast majority of runs clustered within a narrow range. Compared with the Grover algorithm, the distribution width is significantly reduced, and the long-tail effect is almost entirely eliminated, indicating that the algorithm can complete template retrieval with nearly fixed complexity in most cases.

From the perspective of average performance, the mean number of oracle calls for the Long algorithm under the two schemes is also very close. The blue and red dashed lines nearly overlap, indicating that even when only a single quantum counting result is used (Scheme (1)), the algorithm can still maintain a high success probability. This can be understood from the phase-matching mechanism of the Long algorithm. Unlike the fixed \( \pi \)-phase in the standard Grover algorithm, the Long algorithm selects an iteration number \( J_s \) satisfying Eq.~\eqref{equ:LONG_Js_min} and introduces a phase parameter \( \phi \), which ideally enables unit success probability. Moreover, the Long algorithm is less sensitive to estimation errors in \( \theta \), so even when \( \theta_\ast \) has some uncertainty, the deviation in the final rotation angle remains small, preserving a high success probability.

From the perspective of distribution shape, the oracle call counts of the Long algorithm exhibit a concentrated unimodal distribution, with the peak located close to the typical number of calls required for a successful search. This indicates that most simulation runs succeed within one or only a few searches, so the total number of oracle calls is typically confined to a narrow range near the theoretical value. This represents a clear improvement over the Grover algorithm, which often requires multiple repetitions.

Furthermore, the difference between the two schemes is significantly reduced in the Long algorithm. Since the phase-matching mechanism already mitigates the impact of errors in \( \theta \), the performance gain from re-running quantum counting is relatively limited. This further demonstrates the improved tolerance of the Long algorithm to quantum counting errors.

Finally, compared with the computational cost of classical exhaustive search (black dashed line), both quantum algorithms require significantly fewer oracle calls than the classical \( O(N) \) complexity, preserving the quadratic speedup. However, in practical simulations, the Long algorithm not only maintains the same theoretical complexity but also significantly reduces the variance of the complexity distribution, making the runtime more predictable. This concentration of complexity is particularly important in practical quantum computing applications, as it reduces the occurrence of extreme cases and improves overall system stability and resource utilization.

The results of Scheme (2) are shown in Figs.~\ref{fig:sims800_p_vary_g_l} and~\ref{fig:sims870_p_vary_g_l}, which illustrate the impact of varying the quantum counting precision parameter \( p = 10, 11, 12 \) on the performance of the Grover and Long algorithms under \( \rho_{\textrm{thr}} = 800 \) and \( \rho_{\textrm{thr}} = 870 \), respectively. It can be observed that, in these test cases, the effect of quantum counting precision on algorithm efficiency is relatively small, with oracle call counts remaining concentrated within a narrow range. The fundamental reason for the sensitivity of the Grover algorithm to both the threshold \( \rho_{\textrm{thr}} \) and the precision \( p \) lies in its sensitivity to errors in the estimation of the angle \( \theta \) between the initial quantum state and the non-target subspace. In contrast, the Long algorithm alleviates this sensitivity through the phase-matching condition, thereby enhancing its robustness.

In addition, for the threshold sensitivity observed when reproducing the results of Ref.~\cite{gao2022quantum} for the GW150914 signal, further tests were conducted, with the comparison results shown in Fig.~\ref{fig:gw150914_g_l}. These results further demonstrate the robustness of the Long algorithm.

It should be emphasized that the above comparison primarily uses the number of oracle function \( f \) calls as a measure of algorithmic complexity. While this metric captures the dominant cost in the theoretical model, it does not fully reflect the total quantum resource consumption in practical implementations on quantum hardware. Specifically, the Long algorithm introduces a tunable phase factor \( \phi \) to achieve phase matching, improving success probability and reducing sensitivity to estimation errors. While this enhances robustness at the algorithmic level, it also implies a more complex quantum circuit structure compared to the standard Grover algorithm. In the Grover algorithm, the oracle and diffusion operators typically involve fixed \( \pi \)-phase inversions, whereas the Long algorithm requires controlled phase operations with specific phase \( \phi \), or equivalent rotation gates. These operations generally need to be decomposed into sequences of elementary quantum gates, thereby increasing circuit depth and gate count. Therefore, although the Long algorithm demonstrates improved stability and average complexity in terms of oracle calls during template retrieval, its overall resource consumption on actual quantum hardware should be carefully evaluated by considering factors such as circuit depth, gate count, precision of controlled phase operations, and the accumulation of quantum errors.

\textit{Conclusion\textemdash}In this work, by adopting the Long algorithm, a modified version of Grover’s algorithm, we propose a quantum search framework that preserves the optimal $O(\sqrt{N})$ complexity of Grover-type algorithms while exhibiting enhanced robustness compare to , thereby improving the search performance and usability.

\textit{Acknowledgments\textemdash}We thank Prof. Gui-Lu Long for stimulating discussions on the Long algorithm. 
This research is supported by the Research Funds of
Hangzhou Institute for Advanced Study, UCAS, and partly
funded by the Strategic Priority Research Program of the
Chinese Academy of Sciences under Grant
No. XDA15021100, and the Fundamental Research
Funds for the Central Universities.


\bibliography{ref}

@article{durr1996quantum,
  title={A quantum algorithm for finding the minimum},
  author={Durr, Christoph and Hoyer, Peter},
  journal={arXiv preprint quant-ph/9607014},
  year={1996}
}

@article{wei2020quantum,
  title={Quantum algorithms for jet clustering},
  author={Wei, Annie Y and Naik, Preksha and Harrow, Aram W and Thaler, Jesse},
  journal={Physical Review D},
  volume={101},
  number={9},
  pages={094015},
  year={2020},
  publisher={APS}
}

@article{tezuka2022grover,
  title={Grover search revisited: Application to image pattern matching},
  author={Tezuka, Hiroyuki and Nakaji, Kouhei and Satoh, Takahiko and Yamamoto, Naoki},
  journal={Physical Review A},
  volume={105},
  number={3},
  pages={032440},
  year={2022},
  publisher={APS}
}

@article{somiya2012detector,
  title={Detector configuration of KAGRA--the Japanese cryogenic gravitational-wave detector},
  author={Somiya, Kentaro and KAGRA collaboration and others},
  journal={Classical and Quantum Gravity},
  volume={29},
  number={12},
  pages={124007},
  year={2012},
  publisher={IOP Publishing}
}

@article{aasi2015advanced,
  title={Advanced ligo},
  author={Aasi, Junaid and Abbott, BP and Abbott, Richard and Abbott, Thomas and Abernathy, MR and Ackley, Kendall and Adams, Carl and Adams, Thomas and Addesso, Paolo and Adhikari, RX and others},
  journal={Classical and quantum gravity},
  volume={32},
  number={7},
  pages={074001},
  year={2015},
  publisher={IOP Publishing}
}

@article{acernese2014advanced,
  title={Advanced Virgo: a second-generation interferometric gravitational wave detector},
  author={Acernese, Fausto and Agathos, M and Agatsuma, K and Aisa, Damiano and Allemandou, N and Allocca, Aea and Amarni, J and Astone, Pia and Balestri, G and Ballardin, G and others},
  journal={Classical and Quantum Gravity},
  volume={32},
  number={2},
  pages={024001},
  year={2014},
  publisher={IOP Publishing}
}

@article{gao2022quantum,
	title={Quantum algorithm for gravitational-wave matched filtering},
	author={Gao, Sijia and Hayes, Fergus and Croke, Sarah and Messenger, Chris and Veitch, John},
	journal={Physical Review Research},
	volume={4},
	number={2},
	pages={023006},
	year={2022},
    publisher = {American Physical Society},
    doi = {10.1103/PhysRevResearch.4.023006},
    url = {https://link.aps.org/doi/10.1103/PhysRevResearch.4.023006}
}

@article{guo2025quantum,
  title={Quantum search for gravitational wave of massive black hole binaries},
  author={Guo, Fangzhou and He, Jibo},
  journal={Physical Review D},
  volume={112},
  number={8},
  pages={083004},
  year={2025},
  publisher={APS}
}

@article{abbott2016observation,
	title={Observation of gravitational waves from a binary black hole merger},
	author={Abbott, Benjamin P and Abbott, Richard and Abbott, Thomas D and Abernathy, Matthew R and Acernese, Fausto and Ackley, Kendall and Adams, Carl and Adams, Thomas and Addesso, Paolo and Adhikari, Rana X and others},
    journal = {Phys. Rev. Lett.},
    volume = {116},
    pages = {061102},
    year = {2016},
    publisher = {American Physical Society},
    doi = {10.1103/PhysRevLett.116.061102},
    url ={https://link.aps.org/doi/10.1103/PhysRevLett.116.061102}

}

@misc{amaro2017laser,
      title={Laser Interferometer Space Antenna}, 
      author={Amaro-Seoane, Pau and Audley, Heather and Babak, Stanislav and Baker, John and Barausse, Enrico and Bender, Peter and Berti, Emanuele and Binetruy, Pierre and Born, Michael and Bortoluzzi, Daniele and others},
      year={2017},
      eprint={1702.00786},
      archivePrefix={arXiv},
      primaryClass={astro-ph.IM},
      url={https://arxiv.org/abs/1702.00786}, 
}

@article{hu2017taiji,
    author = {Hu, Wen-Rui and Wu, Yue-Liang},
    title = {The Taiji Program in Space for gravitational wave physics and the nature of gravity},
    journal = {National Science Review},
    volume = {4},
    number = {5},
    pages = {685-686},
    year = {2017},
    month = {10},
    issn = {2095-5138},
    doi = {10.1093/nsr/nwx116},
    url = {https://doi.org/10.1093/nsr/nwx116},
}

@article{luo2016tianqin,
  title = {TianQin: a space-borne gravitational wave detector},
  author={Luo, Jun and Chen, Li-Sheng and Duan, Hui-Zong and Gong, Yun-Gui and Hu, Shoucun and Ji, Jianghui and Liu, Qi and Mei, Jianwei and Milyukov, Vadim and Sazhin, Mikhail and others},
  journal = {Classical and Quantum Gravity},
  volume = {33},
  number = {3},
  pages = {035010},
  year = {2016},
  month = {jan},
  publisher = {IOP Publishing},
  doi = {10.1088/0264-9381/33/3/035010},
  url = {https://dx.doi.org/10.1088/0264-9381/33/3/035010},
}

@article{punturo2010einstein,
  title={The Einstein Telescope: A third-generation gravitational wave observatory},
  author={Punturo, M and Abernathy, Matt and Acernese, Fausto and Allen, Bruce and Andersson, Nils and Arun, K and Barone, Fabrizio and Barr, Bryan and Barsuglia, Matteo and Beker, Mark and others},
  journal={Classical and Quantum Gravity},
  volume={27},
  number={19},
  pages={194002},
  year={2010},
  publisher={IOP Publishing}
}

@article{reitze2019cosmic,
  title={Cosmic explorer: the US contribution to gravitational-wave astronomy beyond LIGO},
  author={Reitze, David and Adhikari, Rana X and Ballmer, Stefan and Barish, Barry and Barsotti, Lisa and Billingsley, GariLynn and Brown, Duncan A and Chen, Yanbei and Coyne, Dennis and Eisenstein, Robert and others},
  journal={arXiv preprint arXiv:1907.04833},
  year={2019}
}

@article{owen1999matched,
  title={Matched filtering of gravitational waves from inspiraling compact binaries: Computational cost and template placement},
  author={Owen, Benjamin J and Sathyaprakash, Bangalore Suryanarayana},
  journal = {Phys. Rev. D},
  volume={60},
  number={2},
  pages={022002},
  year={1999},
  publisher = {American Physical Society},
  doi = {10.1103/PhysRevD.60.022002},
  url = {https://link.aps.org/doi/10.1103/PhysRevD.60.022002}
}

@inproceedings{grover1996fast,
	title={A fast quantum mechanical algorithm for database search},
	author={Grover, Lov K},
	booktitle={Proceedings of the twenty-eighth annual ACM symposium on Theory of computing},
	pages={212-219},
    numpages = {8},
	year={1996},
    isbn = {0897917855},
    publisher = {Association for Computing Machinery},
    url = {https://doi.org/10.1145/237814.237866},
    doi = {10.1145/237814.237866},
}

@misc{pythoncode,
  accessed = {2025-05-13},
  author   = {Fergus Hayes},
  title    = {quantum matched filter},
  note = {\url{https://github.com/Fergus-Hayes/quantum-matched-filter}}
}

@article{long2001grover,
  title={Grover algorithm with zero theoretical failure rate},
  author={Long, Gui-Lu},
  journal={Physical Review A},
  volume={64},
  number={2},
  pages={022307},
  year={2001},
  publisher={APS},
  doi = {10.1103/PhysRevA.64.022307},
  url = {https://link.aps.org/doi/10.1103/PhysRevA.64.022307}
}

@article{katz2020gpu,
  title={GPU-accelerated massive black hole binary parameter estimation with LISA},
  author={Katz, Michael L and Marsat, Sylvain and Chua, Alvin JK and Babak, Stanislav and Larson, Shane L},
  journal={Physical Review D},
  volume={102},
  number={2},
  pages={023033},
  year={2020},
  publisher={APS},
  doi = {10.1103/PhysRevD.110.103026},
  url = {https://link.aps.org/doi/10.1103/PhysRevD.110.103026}
}

@article{katz2022fully,
  title={Fully automated end-to-end pipeline for massive black hole binary signal extraction from LISA data},
  author={Katz, Michael L},
  journal={Physical Review D},
  volume={105},
  number={4},
  pages={044055},
  year={2022},
  publisher={APS},
  doi = {10.1103/PhysRevD.105.044055},
  url = {https://link.aps.org/doi/10.1103/PhysRevD.105.044055}
}

@software{michael_katz_2021_5730688,
  author       = {Michael Katz},
  title        = {mikekatz04/BBHx: First official public release},
  month        = nov,
  year         = 2021,
  publisher    = {Zenodo},
  version      = {v1.0.0},
  doi          = {10.5281/zenodo.5730688},
  url          = {https://doi.org/10.5281/zenodo.5730688},
}

@article{zhao2012geometric,
  title={Geometric pictures for quantum search algorithms},
  author={Zhao, Lian-Jie and Li, Yan-Song and Hao, Liang and Zhou, Tao and Long, Gui Lu},
  journal={Quantum Information Processing},
  volume={11},
  number={2},
  pages={325--340},
  year={2012},
  publisher={Springer}
}

\end{document}